\documentclass[11pt]{article}

\usepackage{arxiv}

\usepackage[utf8]{inputenc} 
\usepackage[T1]{fontenc}
\usepackage{amsmath,amsfonts,amssymb,bm}
\usepackage{graphicx,color}
\usepackage{setspace}
\doublespace   
\usepackage{fancyhdr}
\usepackage[absolute,overlay]{textpos}
\usepackage{multicol}
\usepackage[authoryear,round]{natbib}
\usepackage{booktabs} 
\usepackage{caption}
\usepackage{subcaption}
\usepackage{url}
\usepackage{float}
\usepackage{multirow}
\usepackage{rotating}
\usepackage{hhline}
\usepackage{algorithm}
\usepackage{algpseudocode}

\newcommand{\tCollapsedurl}{\if0\blind
	{
		\url{\tCollapsedurl}
	} \fi
	\if1\blind
	{ 
		\url{github.com/}\texttt{[url redacted in blinded version]}
	} \fi}

\title{Does wind affect the orientation of vegetation stripes? A copula-based mixture model for axial and circular data}

\author{
Marco Mingione\\
    \scriptsize{Dpt. of Political Sciences}\\
    \scriptsize{Roma Tre University}\\
    \scriptsize{\texttt{marco.mingione@uniroma3.it}}\\
    \And
Francesco Lagona\\
    \scriptsize{Dpt. of Political Sciences}\\
    \scriptsize{Roma Tre University}
\And
    Priyanka Nagar\\
    \scriptsize{Dpt of Statistics and Actuarial Science}\\
    \scriptsize{Stellenbosch University}
\And
Francois von Holtzhausen\\
    \scriptsize{Dpt. of Statistics}\\
    \scriptsize{University of Pretoria}
\And
Andriette Bekker\\
    \scriptsize{Centre of Environmental Sciences and Dpt. of Statistics}\\
    \scriptsize{University of Pretoria}
\And
Janine Schoombie\\
    \scriptsize{Dpt. of Mechanical and Aeronautical Engineering}\\
    \scriptsize{University of Pretoria}
\And
Peter C. le Roux\\
    \scriptsize{Dpt. of Plant and Soil Sciences}\\
    \scriptsize{University of Pretoria}
}

\begin{document}
\maketitle
\begin{abstract}
Motivated by a case study of vegetation patterns, we introduce a mixture model with concomitant variables to examine the association between the orientation of vegetation stripes and wind direction. The proposal relies on a novel copula-based bivariate distribution for mixed axial and circular observations and provides a parsimonious and computationally tractable approach to examine the dependence of two environmental variables observed in a complex manifold. The findings suggest that dominant winds shape the orientation of vegetation stripes through a mechanism of neighbouring plants providing wind shelter to downwind individuals.
\end{abstract}

\section{Introduction}
\label{sec:introduction}
Vegetation patchiness is frequently observed in climatically extreme environments, with linear vegetation features (i.e. vegetation stripes or bands) recorded in a diversity of arid, polar, alpine and wetland habitats \citep{deblauwe2008global, bekker2008linear, rietkerk2004self}. However, there is still an incomplete understanding of the environmental factors that influence the occurrence and nature of vegetation patterning \citep{alftine2004directional}. For example, while wind direction has been hypothesized to drive the orientation of vegetation stripes \citep[with support from simulation studies; e.g.][]{zhang2017vegetation}, no field studies have quantitatively tested this idea beyond small scales or single landforms \citep{baartman2018effect, morgan2010wind}. 

The statistical analysis of these data is inherently different from the traditional analysis of bivariate continuous data. Specifically, while wind directions are observed in the form of circular data, the orientations of vegetation stripes are axial data, that is, random segments in which neither end can be identified as the starting point. The literature of directional statistics offers several options to model circular and axial data separately \citep{wells2010advances,Pewsey_etal2021,Ley_etal2017}. To the best of the authors' knowledge, however, methods for the analysis of the joint distribution of mixed axial and circular data have been overlooked. 

Here, we propose a novel class of bivariate distributions for mixed axial and circular data using a copula-based approach. Copulas allow the construction of new bivariate distributions with the desired marginals and have already been used in directional statistics to specify distributions with circular marginals \citep{Wehrly_etal1980, Jones_etal2015, imoto_abe2021_copula, hodel2022_cilindrical_copulas}. We extend this approach to the mixed axial-circular setting by proposing a new copula that relies on the toroidal density proposed by \citet{Kato_Pewsey2015}.   

Data heterogeneity is a further complication in this study. The joint distribution of vegetation stripe orientation and wind direction may vary according to environmental conditions. We therefore fit the data by a mixture of axial-circular bivariate distributions, where mixing weights depend on the available environmental covariates, therefore exploiting a mixture model with concomitant variables \citep{dayton1988_concvariable_mixture}. By taking this approach, we are capable of segmenting the data according to clusters that feature different degrees of dependence between the outcomes and are simultaneously associated with the levels taken by the observed environmental factors.

The remainder of the paper is organized as follows: Section \ref{sec:data} describes the motivating example and the available data; Section \ref{sec:methods} introduces the novel joint axial-circular distribution by means of a specific copula and illustrates how to embed it into a finite mixture modelling framework; Section \ref{sec:estimation} encompasses the estimation procedure obtained via the EM algorithm; Section \ref{sec:simulation} shows the performances of the proposed model on different simulated scenarios, while Section \ref{sec:results} includes the real-data application. Final points of discussion are summarized in Section \ref{sec:discussion}.
 
\section{Vegetation stripes on Marion island}\label{sec:data}

\begin{figure}[ht]
    \centering
    \begin{subfigure}[b]{.45\textwidth}
        \centering
        \includegraphics[width = .95\textwidth]{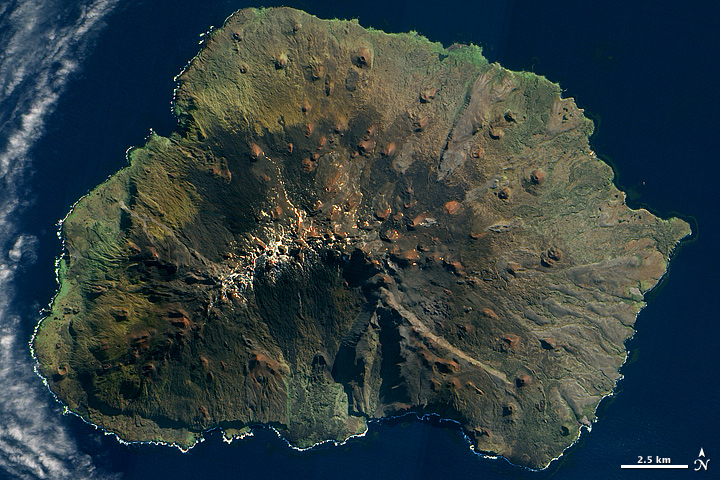}
        \caption{}
        \label{fig:marionisland}
    \end{subfigure}
    \begin{subfigure}[b]{.45\textwidth}
        \centering
        \includegraphics[width = .95\textwidth]{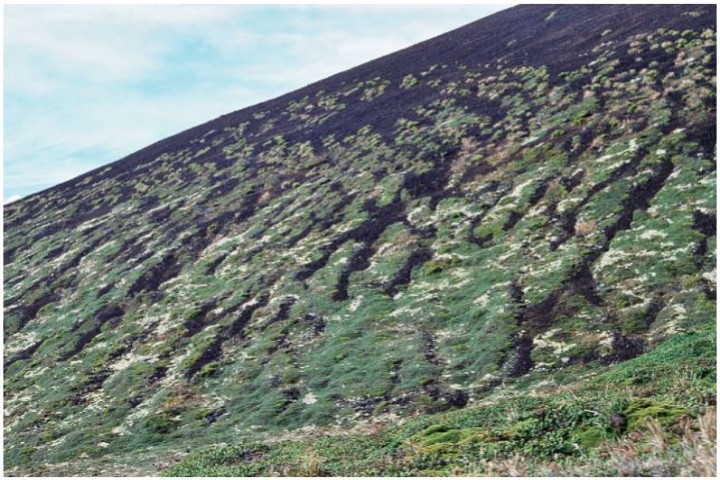}
        \caption{}
        \label{fig:conestripes}
    \end{subfigure}
     \caption{(a) Aerial view of Marion Island and (b) frontal view of a scoria cone with visible vegetation stripe patterns, reprinted from “Vegetation of subantarctic Marion and Prince Edward Islands” by \citeauthor{smith2006vegetation}, \citeyear{smith2006vegetation}, Strelitzia, 19, 708. Copyright 2006.}
\end{figure}

Marion Island (see Figure \ref{fig:marionisland}) is one of the six island archipelagos in the sub-Antarctic, located just north of the Antarctic Polar Convergence in the South Indian Ocean. The island's climate is maritime, characterised by cool temperatures that are very buffered (mean daily temperature range < 3ºC and mean seasonal temperature difference is c. 4ºC) and high year-round precipitation (c. 2000 mm p.a.). The island is a very windy environment, experiencing < 10 windstill days per year between 1992 and 2003, and gale force winds (> 15 m.sec-1) on more than 100 days per year. The strongest (averaging > 10 m.sec-1) and most common winds blow in from the north-west of the island (i.e. 60\% of observation being north-westerly winds), with easterly winds being much less common (<10\% of occurrences) and weaker (<6 m.sec-1). However, Marion Island's climate is changing rapidly, with temperatures rising at double the mean global warming rate, and strong declines in rainfall recorded since the 1960s. Changes in wind conditions over the same time period are less clear, although mean wind speed has increased and wind direction shifted to a more northerly direction (particularly in late summer and autumn). Further details are provided by \citet{chown2008prince, le2008changes}.

 The flora of Marion Island comprises c. 21 indigenous vascular plant species (although a much larger number of lichen, moss and liverwort species also occur on the island), with vegetation cover ranging from complete (i.e. no bare soil) at the coast to an entire lack of vascular plants above 800 m a.s.l. \citep{smith2006vegetation}. At mid- and high-altitudes, vegetation is typically patchy on the island, with many species co-occurring in vegetation patches or stripes initiated by the cushion plant \textit{Azorella selago} \citep{mcgeoch2008spatial,haussmann2009interactions}. 

Vegetation stripes are particularly pronounced on Marion Island's pyroclastic scoria cones (see Figure \ref{fig:conestripes}), of which 130 occur around the island and which were formed during volcanic eruptions.  Scoria cones' surface is comprised of loose, coarse pebble-like rocks (scoria), and is typically disproportionately exposed to winds due to the cones often being considerably taller than the adjacent landforms (e.g. 50 - 200 m elevation) and typically being located away from cliffs and ridges that could otherwise offer some wind shelter. The vegetation on these scoria cones is dominated by four plant species: \textit{Azorella selago}, the grass \textit{Agrostis magellanica}, the mat-forming fern \textit{Blechnum penna-marina}, and the low-growing woody shrub \textit{Acaena magellanica}. These species often grow in approximately linear formations on scoria cones, and have been hypothesized to reflect how neighbouring plants benefit each other by providing wind shelter and protection from burial by scoria in their immediate environment.  While it might be assumed that vegetation stripes would typically be aligned with the slope (i.e. maximizing protection from burial by scoria and dominated by gravity-related processes), it has been observed that the orientation of vegetation stripes varies considerably, both within individual scoria cones (e.g. comparing different sides of a single scoria cone) and across scoria cones. Further details are provided by \citet{le2008spatial}. 

The orientation of other linear features in the sub-Antarctic also shows variation in their orientation. For example, variation in the orientation of sorted stripes (geomorphological features where stripes of stones of different sizes arise from repeated freezing of the soil) has been hypothesized to be due to diverse processes including, e.g., differences in snow cover \citep{hall1994some}. However, of the potential environmental factors affecting these patterns, wind direction has repeatedly been inferred as a key driver of the orientation of sorted stripes, with this feature typically running parallel with dominant winds \citep{hall1979sorted, holness2001orientation,hedding2015aeolian}, suggesting the importance of wind patterns in explaining the directionality of linear features in this system.

\begin{figure}[ht]
    \centering
    \begin{subfigure}[b]{.3\textwidth}
    \includegraphics[width = .99\textwidth]{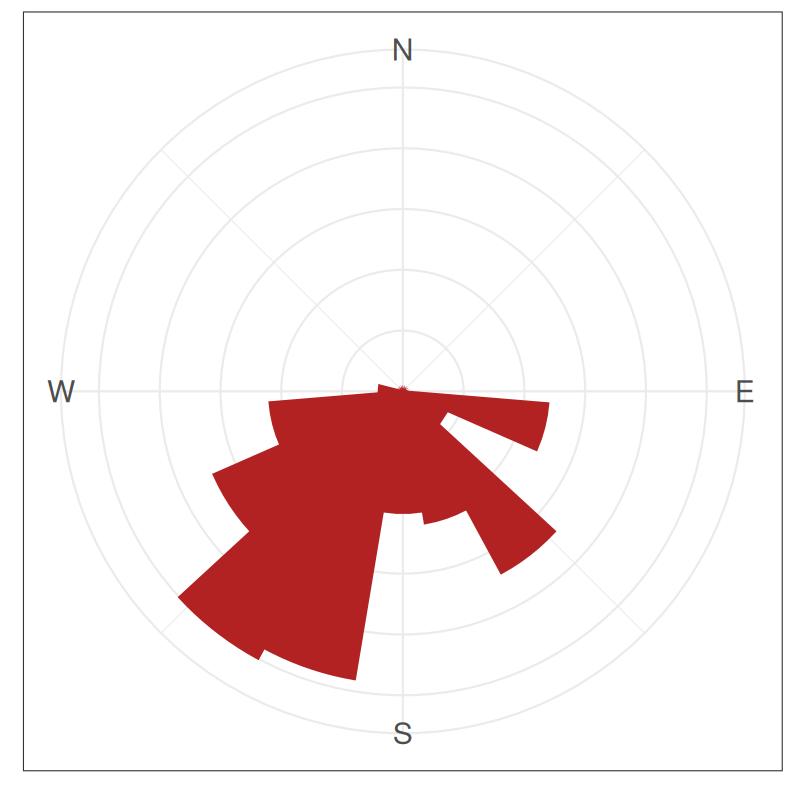}
    \caption{}
    \label{stripedir}
    \end{subfigure}
    \begin{subfigure}[b]{.3\textwidth}
    \includegraphics[width = .99\textwidth]{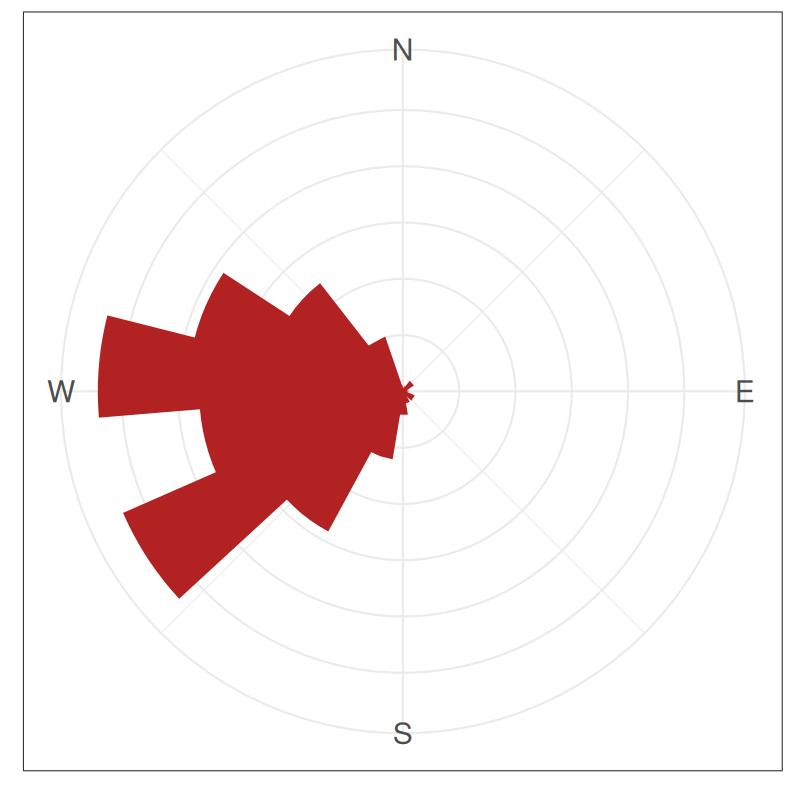}
    \caption{}
    \label{winddir}
    \end{subfigure}
    \begin{subfigure}[b]{.3\textwidth}
    \includegraphics[width = .99\textwidth]{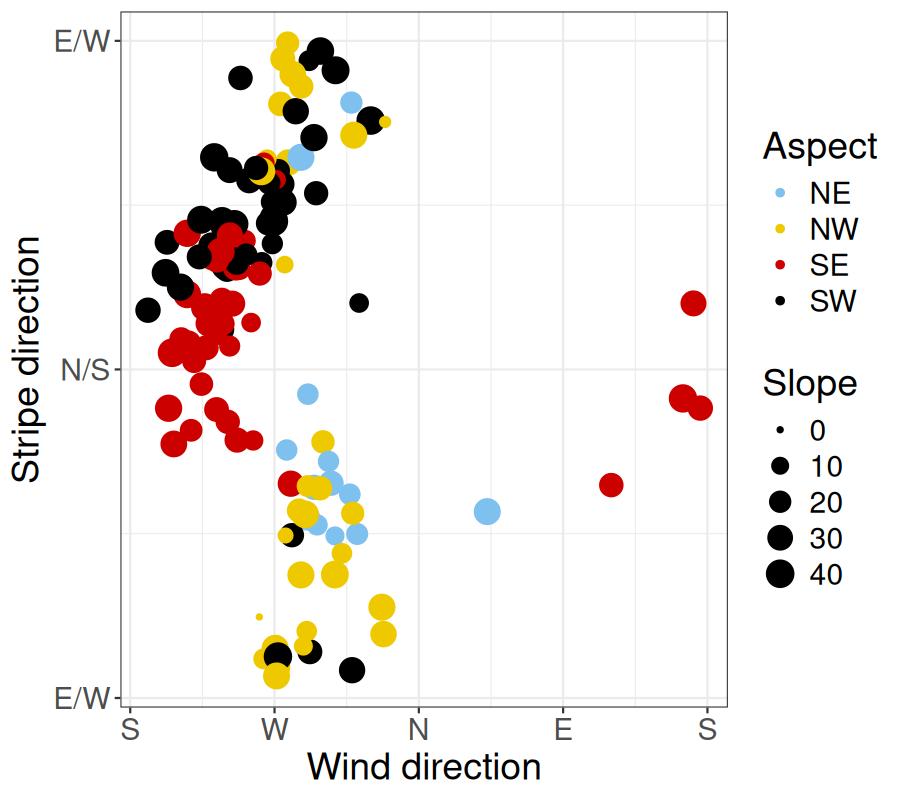}
    \caption{}
    \label{scatter}
    \end{subfigure}
    \caption{Marginal distributions of stripe orientations (a) and wind directions (b), paired with their joint distribution (c). Data points are colored according to the aspect of the pyroclastic scoria cone where the stripe was observed, while their size is proportional to the associated cone's slope (slope indicated in degrees).}
    \label{fig:enter-label}
\end{figure}

In this study we extracted vegetation stripe orientation from 35 scoria cones, measuring vegetation stripes on four different aspects (NE, NW, SE, SW) of each cone where vegetation stripes were present (resulting in data from $n=133$ locations). This subset of the island's scoria cones was selected because (1) they exhibited vegetation stripes on at least one aspect, (2) they had a regular shape, i.e. approximately conical form, and (3) they had a relatively consistent orientation within each aspect. Stripe orientation was extracted from Google Earth imagery, using a circular average of five vegetation stripes per aspect per scoria cone. Stripe aspect (i.e. orientation relative to north) and angle (i.e. steepness/slope) were extracted for each vegetation stripe from a 1 m-resolution digital surface model. We expect that the orientation of vegetation stripes on steeper slopes would be less related to wind direction (i.e., to be more strongly linked to the downslope direction due to stronger gravitational effects) and that the orientation of stripes on aspects more sheltered from dominant winds would have weaker correlations with wind conditions. Figure \ref{stripedir} shows the marginal distribution of stripe orientation. This rose diagram should be interpreted by recalling that vegetation stripes are random segments in which neither end can be identified as the starting point. As a result, for example, a north-eastern orientation is indistinguishable from a south-western orientation. Accordingly, Figure \ref{stripedir} highlights the presence of vegetation stripes that are mostly oriented towards NE/SW, though there is a smaller subset of locations for which the average stripe orientation is NW/SE. 

Due to the strong dominance of westerly winds in the sub-Antarctic region, Marion Island offers an excellent study system for examining the impacts of wind direction on vegetation patterns. Estimated wind speed and direction at the location of each vegetation stripe were extracted from \citet{goddard2022investigation}. While previous investigations of this type have had to assume that wind data from a single weather station on the island's northeastern coast is representative of the entire island, a recently developed Computational Fluid Dynamics (CFD) model has generated accurate estimates for wind patterns at c. 30 m-resolution \citet{goddard2022investigation}, enabling this study to account for local variation in wind speed and direction due to the island's heterogeneous topography. The CFD model was validated using data collected continuously from April 2018 to March 2020 at 17 locations across Marion Island at 0.5 m and 1 m above the ground surface \citep[details in ][]{goddard2022investigation}. Figure \ref{winddir} confirms that westerly winds are predominant in the study area. However, the two modes in the rose diagram suggest heterogeneity in wind conditions due to the island's peculiar topographic features. 

The joint distribution of wind and stripe directions is displayed in Figure \ref{scatter}. The domain of these points is the manifold obtained by the cartesian product of a circle and a semi-circle, and therefore Figure \ref{scatter} should be interpreted as the unwrapped counterpart of such manifold. Here, data points are colored according to the cone's aspect, while their size is proportional to the associated cone's slope. While the latter does not highlight any specific pattern, the cone's aspect offers clear evidence of separated groups: locations experiencing southwesterly winds with stripes oriented towards north/northeast mostly belong to the southern part of the cone; on the other hand, locations on the northern part of the cone mostly experience northwesterly winds and show stripe directions in almost all directions (with north-south orientations being the least common) with respect to the cone shape. Fitting these data by a mixture of bivariate axial-circular distributions with concomitant variables, we are not only able to cluster the data according to different correlation structures but we simultaneously assess the influence of topographic features (e.g., aspect and slope of the cone) on cluster membership.

\section{A mixture model for axial and circular data}\label{sec:methods}
The proposed model is a mixture of bivariate distributions, obtained by binding a univariate axial distribution and a univariate circular distribution through a copula density. In the following, we first introduce the proposed bivariate distribution and, second, we describe the mixture model that we exploit for data analysis.

\subsection{A bivariate circular-axial distribution}\label{subsec:copulaprop}
 
A direction observed in the plane, like wind direction, can be represented by a random angle, often referred to as a continuous circular random variable $X$. Since angles $x$ and $x+2\pi$ are indistinguishable, the probability density function (pdf) of a circular random variable $f_{\rm{circ}}(x)$ is a nonnegative function that integrates to 1 over any interval of length $2\pi$ and it is $2\pi$-periodic, that is, $f_{\rm{circ}}(x+2k\pi)=f_{\rm{circ}}(x)$ for any integer $k$. Due to periodicity, the range of $X$ can be conventionally defined as the interval $[0,2\pi)$, where $0$ is the angle chosen in the unit circle as the zero direction, and the cumulative distribution function (cdf) can be defined as $F_{\rm{circ}}(x)=\int_{0}^{x}f_{\rm{circ}}(t)dt$,  $x \in [0, 2\pi)$. The inverse cdf $F^{-1}_{\rm{circ}}(u)$, $0 \leq u <1$, is accordingly a continuous mapping from the unit interval $[0,1)$ to $[0,2\pi)$. 

\begin{figure}[ht]
    \centering
    \begin{subfigure}[b]{.48\textwidth}
    \includegraphics[width = .99\textwidth]{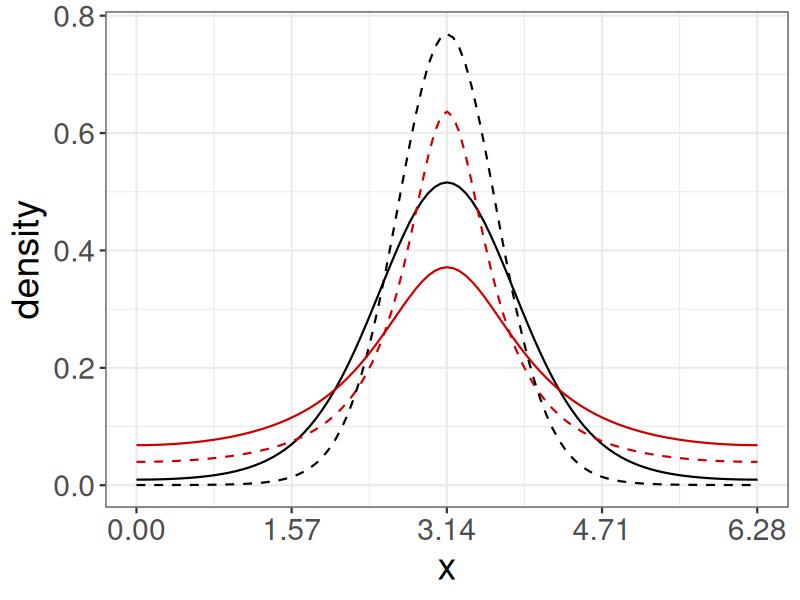}
    \caption{}
    \label{margcirc}
    \end{subfigure}
    \begin{subfigure}[b]{.48\textwidth}
    \includegraphics[width = .99\textwidth]{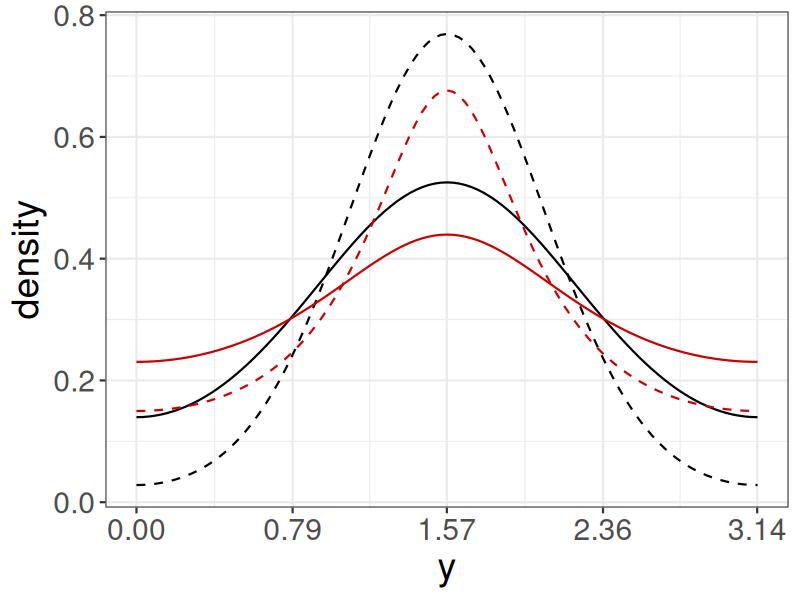}
    \caption{}
    \label{margax}
    \end{subfigure}
     \caption{(a) von Mises (black) and wrapped Cauchy (red) circular distributions, centered at $\mu=\pi$ and concentrations respectively fixed at  $\kappa=2,0.4$ (continuous) and $\kappa=4,0.6$ (dashed). (b) the axial distributions obtained by wrapping the left-hand side distributions on the semi-circle.}
    \label{fig:marginalex}
    \end{figure}

The first two rows of Table~\ref{tab:univariate_densities} display the two most popular examples of parametric circular pdfs, namely the von Mises and the wrapped Cauchy, both symmetric about a location parameter, $\mu \in [0, 2\pi)$, and shaped by a parameter $\kappa$ that regulates the concentration of the distribution around $\mu$. Specifically, when $\kappa=0$, both densities reduce to the circular uniform density $f(x)=1/ (2\pi)$ with cdf $F(x)=x/(2\pi)$ and inverse cdf $F^{-1}(u)=2\pi u$. Figure \ref{margcirc} shows examples of the two densities, centered at $\mu=\pi$ and obtained by considering two values of the concentration parameter. 

\begin{table}[ht]
    \centering
    \large
    \caption{Examples of circular and axial densities.}
    \label{tab:univariate_densities} 
    \begin{tabular}{llccc}
    \toprule
    Name   & Range & Density & Location & Concentration\\ \midrule
         Circular von Mises  & $[0, 2\pi)$&$\frac{\exp(\kappa \cos(x-\mu))}{2\pi I_0(\kappa)}$ & $0 \leq \mu < 2\pi$ & $\kappa \geq 0$\\ \\
         Circular wrapped Cauchy & $[0, 2\pi)$ & $\frac{1-\kappa^2}{2\pi(1+\kappa^2 - 2\kappa\cos(x-\mu))}$ &$0 \leq \mu < 2\pi$ & $0 \leq \kappa <1$\\ \\
         Axial von Mises & $[0, \pi)$ &$\frac{\exp(\kappa \cosh(y-\mu))}{\pi I_0(\kappa)}$  &$0 \leq \mu < \pi$&$\kappa \geq 0$\\ \\
         Axial wrapped Cauchy &  $[0, \pi)$& $\frac{1}{\pi}\frac{1-\kappa^4}{1+\kappa^4-2\kappa^2\cos(2(y-\mu))}$ & $0 \leq \mu < \pi$ &$0 \leq \kappa <1$ \\
         \bottomrule
    \end{tabular}
    
\end{table}

An axial observation, like the orientation of a vegetation stripe, indicates instead the orientation $y$ of a segment in the plane, in which neither end can be identified as the starting point. In other words, axial angles $y$ and $y +\pi$ are indistinguishable. As a result, the pdf $f_{\rm{axial}}(y)$ of a continuous axial random variable $Y$ is any nonnegative function that integrates to 1 over any interval of length $\pi$ and it is $\pi$-periodic,  $f_{\rm{axial}}(y+\pi) = f_{\rm{axial}}(y)$. The interval $[0,\pi)$ is therefore the conventional range of $Y$. Accordingly, the cdf   $F_{\rm{axial}}(y) = \int_{0}^{y}f_{\rm{axial}}(t)dt$, $ 0 \leq y < \pi$ maps the semi-circle to the unit interval, while the inverse cdf $F_{\rm{axial}}^{-1}(v)$, $0 \leq v <1$ maps the unit interval back to the semi-circle. 

Several proposals of axial pdfs have been proposed in the literature \citep{yedlapalli2023toward,iftikhar2022half,abuzaid2018half}. \citet{Arnold_Sengupta2006} notice that if $X$ is a circular random variable, then $Y = X \pmod \pi $ is an axial random variable and obtain axial pdfs by wrapping a circular pdf on $[0,\pi)$. The last two rows of Table \ref{tab:univariate_densities} display the two axial pdfs that can be obtained by wrapping a von Mises and a wrapped Cauchy. Both axial densities are symmetric about a location parameter $\mu \in [0, \pi)$ and shaped by a parameter $\kappa$ that regulates the concentration of the distribution around $\mu$. When $\kappa=0$ both densities reduce to the axial uniform density $f(x)=1/\pi$ with cdf $F(x)=x/\pi$ and inverse cdf $F^{-1}(v)=\pi v$. Figure \ref{margax} shows the axial densities obtained by wrapping a circular von Mises and a circular wrapped Cauchy around the semi-circle.

\begin{figure}[ht]
    \centering
    \begin{subfigure}[b]{.3\textwidth}
    \includegraphics[width = .99\textwidth]{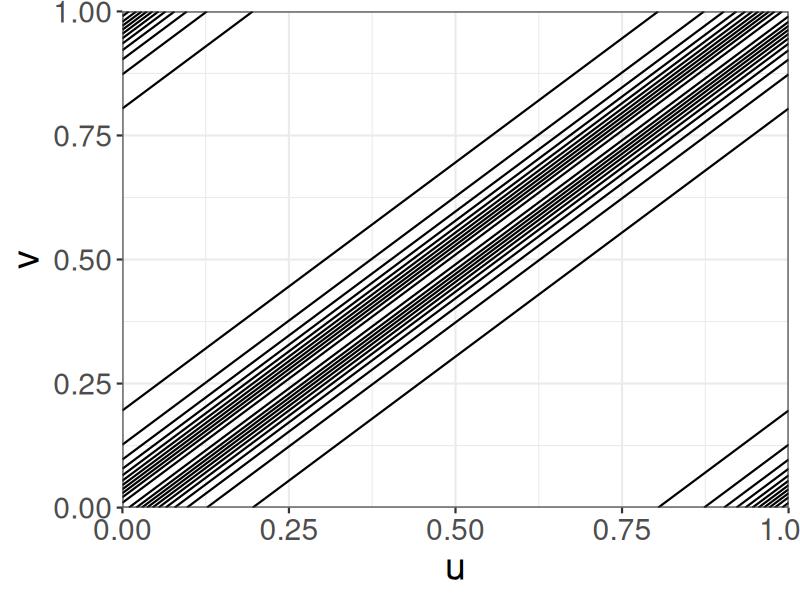}
    \caption{$\rho = 0.7$}
    \end{subfigure}
    \begin{subfigure}[b]{.3\textwidth}
    \includegraphics[width = .99\textwidth]{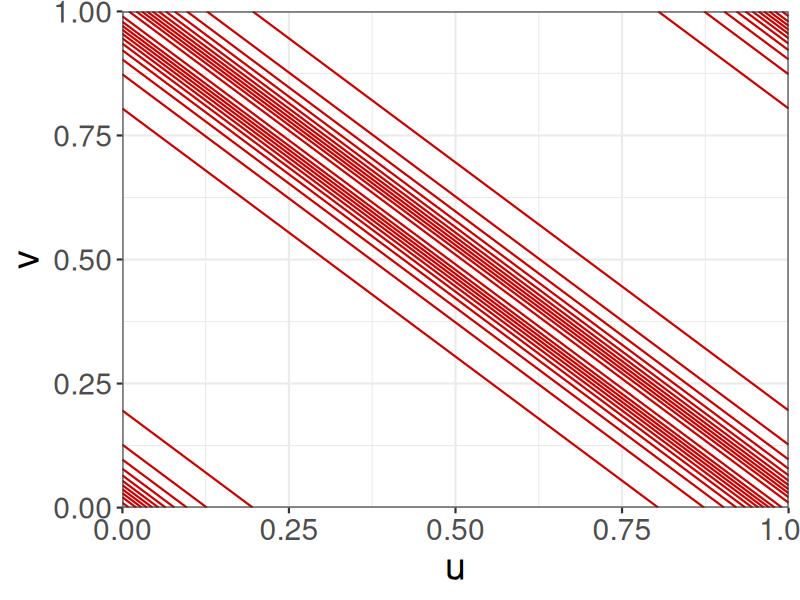}
    \caption{$\rho = -0.7$}
    \end{subfigure}
    \begin{subfigure}[b]{.3\textwidth}
    \includegraphics[width = .99\textwidth]{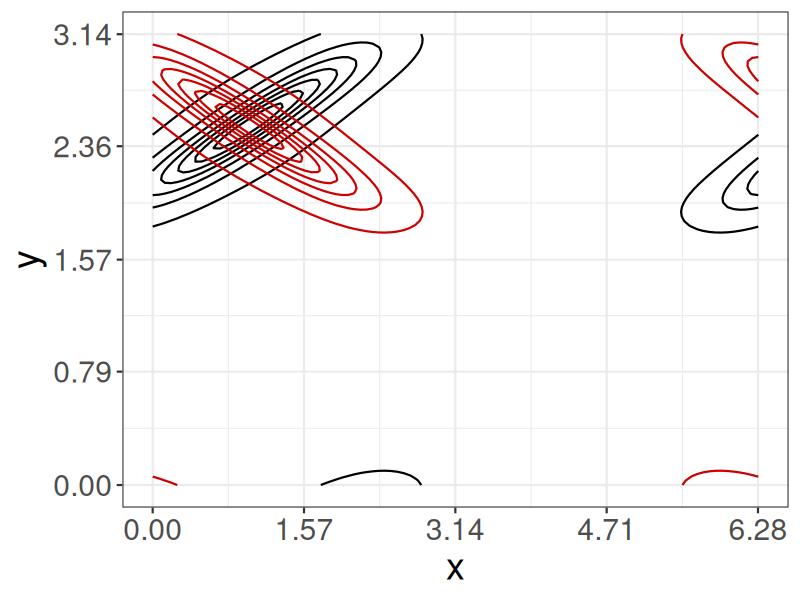}
    \caption{}
    \end{subfigure}
    \begin{subfigure}[b]{.3\textwidth}
    \includegraphics[width = .99\textwidth]{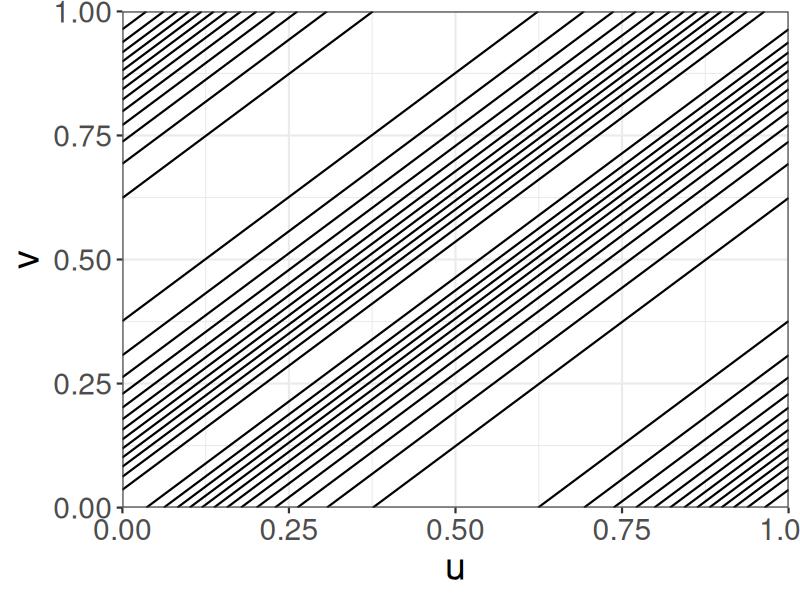}
    \caption{$\rho = 0.3$}
    \end{subfigure}
    \begin{subfigure}[b]{.3\textwidth}
    \includegraphics[width = .99\textwidth]{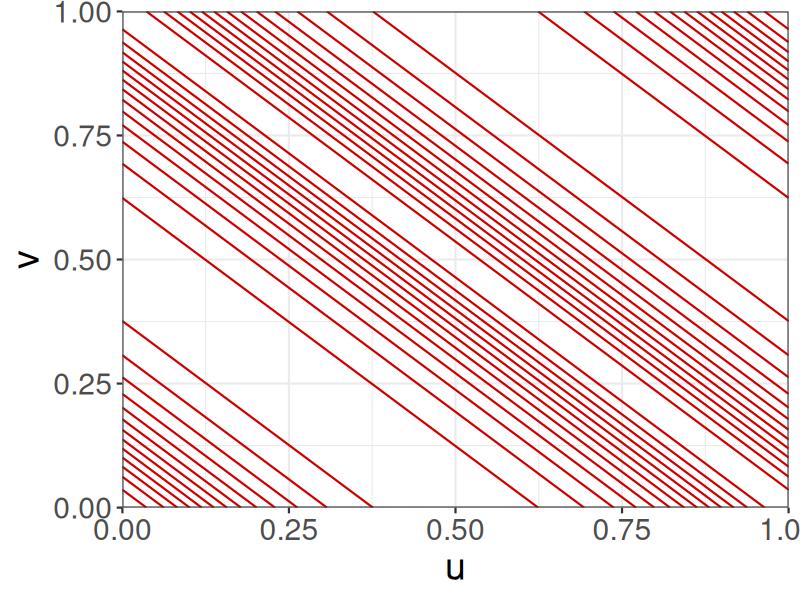}
    \caption{$\rho = -0.3$}
    \end{subfigure}
    \begin{subfigure}[b]{.3\textwidth}
    \includegraphics[width = .99\textwidth]{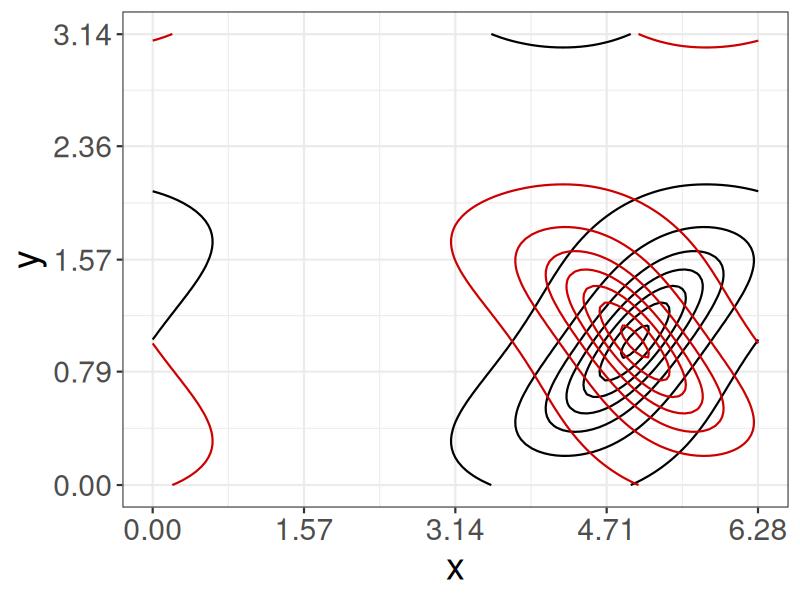}
    \caption{}
    \end{subfigure}
    \caption{Contours of periodic copula densities (circulas) indexed by positive (left panels) and negative (mid panels) correlation parameters and induced bivariate distributions with circular and axial marginals (right panels). Top: circular and axial von Mises marginals with $\mu_{circ} = 1,\kappa_{circ} = 1$ and $\mu_{ax} = 2.5, \kappa_{ax} = 4$. Bottom: circular wrapped Cauchy and axial von Mises marginals with $\mu_{circ} = 5,\kappa_{circ} = 0.4$ and $\mu_{ax} = 1, \kappa_{ax} = 2$. } 
    \label{fig:copulaex}
\end{figure}

Univariate circular and axial cdfs (and their inverses) can be also exploited to specify bivariate distributions with mixed circular and axial marginals, by relying on copulas. Bivariate copulas are bivariate cdfs $C(u,v)$, $0 \leq u,v\leq 1$, with uniform marginals. A popular example is provided by bivariate Gaussian copulas. Let  $\Phi_\rho$ be the cdf of the bivariate standard normal with correlation $\rho$ and marginal cdfs $\Phi$. The Gaussian copula is defined as 
\begin{equation}\label{eq:gaussian_copula}
C_\rho(u,v)=\Phi_\rho(\Phi^{-1}(u), \Phi^{-1}(v)), \qquad 0 \leq u, v \leq 1.\end{equation}
If a bivariate random variable $(X,Y)$ with marginal cdfs $F(x)$ and $F(y)$ admits a density function $f(x,y)$ with respect to the Lebesgue measure on $\mathbb{R^2}$, such density can be written as 
\begin{equation}\label{eq:bivariate density}
    f(x,y)=c_{\rho}(F(x), F(y))f(x)f(y),
\end{equation}   
where $f(x)$ and $f(y)$ are the two marginal densities, obtained by differentiating $F(x)$ and $F(y)$, and $c_\rho$ is the copula density, obtained by differentiating $C_\rho$, namely
\begin{equation}\label{eq:gaussian copula density}
c_{\rho}(F(x), F(y))=\frac{\varphi_\rho(\Phi^{-1}(F(x)),\Phi^{-1}(F(y)))}{\varphi(\Phi^{-1}(F(x)))\varphi(\Phi^{-1}(F(y)))},\end{equation}
where $\varphi_\rho$ is the density of a bivariate standard normal with marginal densities $\varphi$. Equation \eqref{eq:bivariate density} shows that Gaussian copulas can be exploited to specify new bivariate distributions on $\mathbb{R}^2$ with the desired marginals. Unfortunately, though, Gaussian copula densities lack the periodicity requirement that is needed to specify a bivariate distribution with circular marginals, i.e. 
\begin{align}
c_\rho(u,0)=&c_{\rho}(u,1), 0 \leq u \leq 1 \nonumber\\
c_\rho(0,v)=&c_\rho(1,v), 0 \leq v \leq 1 \label{eq:circula_periodicity}.
\end{align}
Gaussian copulas are therefore not suitable for specifying circular distributions. Copula densities that meet the condition in Equation  \eqref{eq:circula_periodicity} are referred to as circula and can be obtained in several ways \citep{garcia_portugues2013,Jones_etal2015, hodel2022_cilindrical_copulas}. Here, we obtain a circula by replacing the bivariate normal density in equation \eqref{eq:gaussian copula density} with a uniparametric bivariate wrapped Cauchy density \citep{Kato_Pewsey2015}
\begin{equation}\label{eq:kato_pewsey}
    g_\rho(x,y) = \frac{1-\rho^2}{4\pi^2\left(1+\rho^2-2\mid \rho\mid \cos(x)\cos(y)-2\rho\sin(x)\sin(y)\right)}    \qquad 0 \leq x,y, < 2\pi , \quad -1\leq \rho \leq 1.
\end{equation}
This density is defined on the cartesian product of two copies of the unit circle (a torus) and, similarly to the standard normal density, it is a unimodal distribution that depends on one correlation parameter $\rho \in [-1,1]$. The univariate marginals of \eqref{eq:kato_pewsey} are uniform densities on the circle, that is $g(x)=g(y)=(2\pi)^{-1}$, $0 \leq x,y < 2\pi$, with cdfs $G(x)=x/(2\pi)$ and $G(y)=y/( 2\pi)$ and inverses $G^{-1}(u)=2\pi u$ and $G^{-1}(v)=2\pi v$. As a result, the proposed circula takes the form   
\begin{align}
c_{\rho}(F(x), F(y))=&\frac{g_\rho\left(G^{-1}(F(x)),G^{-1}(F(y))\right)}{g(G^{-1}( F(x)))g(G^{-1}(F(y))) } \nonumber\\
=&\frac{1-\rho^2}{4\pi^2\left(1+\rho^2-2\mid \rho\mid \cos(2\pi F(x))\cos(2\pi F(y))-2\rho\sin(2\pi F(x))\sin(2\pi F(y))
\right) \frac{1}{4\pi^2} } \nonumber\\
=&\frac{1-\rho^2}{\left(1+\rho^2-2\mid \rho\mid \cos(2\pi F(x))\cos(2\pi F(y))-2\rho\sin(2\pi F(x))\sin(2\pi F(y))
\right) }.
\end{align}
Such circula fulfills conditions \eqref{eq:circula_periodicity} and can be therefore exploited to specify a mixed circular-axial bivariate density
\begin{equation}\label{eq:circ_axial_density}
f(x,y)=c_\rho(F_{\rm circ}(x),F_{\rm axial}(y))f_{\rm circ}(x)f_{\rm axial}(y), \qquad 0 \leq x <2\pi, \quad 0 \leq y < \pi\end{equation}
 with the desired marginal pdfs $f_{\rm circ}(x)$ and $f_{\rm axial}(y)$. If $\rho=0$, then $c_\rho$ is identically equal to 1 and the proposed density \eqref{eq:circ_axial_density} reduces to the product density of two independent circular and axial densities. Otherwise, as $\rho$ increases toward either 1 or -1, $c_\rho$ accumulates the highest values in the neighborhood of the diagonals and the corner points of the unit square and, as a result, the bivariate density \eqref{eq:circ_axial_density} appears more concentrated around a line that is wrapped around the support $[0, 2\pi) \times [0,\pi)$. Figure \ref{fig:copulaex} shows the contour plots of both the circular $c_\rho$ and the bivariate axial-circular density \eqref{eq:circ_axial_density}, for a battery of values of the correlation parameter $\rho$.

\subsection{A mixture model with circular and axial components}\label{subsec:mixture}
Though the definition of an axial-circular distribution represents a novelty per se, the exploratory analysis of the available data highlighted the presence of latent heterogeneity and multimodality, especially for the stripe direction. This heterogeneity likely depends on biotic and abiotic factors, such as soil moisture gradients, substrate temperature and microtopographic variations, which collectively influence the orientation and formation of these patterns. A single distribution could not adequately capture this differential behaviour, and therefore we embed the proposed density in Section \ref{subsec:copulaprop} within a parametric mixture model setting.  

Specifically, we assume that a pair $(x,y)$ of mixed circular and axial observation is a sample drawn from a mixture of $J$ parametric axial-circular densities
\begin{equation}\label{eq:mixture1}
f(x,y) = \sum_{j=1}^{J}\pi_j f(x,y; \bm{\theta}_j),
\end{equation}
where $\pi_1, \ldots, \pi_J$ are multinomial probabilities that sum to 1 and 
\begin{equation}\label{eq:parametric density}
f(x,y; \bm{\theta}_j)=c_{\rho_j}(F(x, \bm{\gamma_j}),F(y;\bm{\alpha}_j))f(x;\bm{\gamma}_j)f(y;\bm{\alpha}_j)
\qquad \bm{\theta}_j=(\bm{\gamma}_j, \bm{\alpha}_j, \rho_j)
\end{equation}
is the circular-axial density in \eqref{eq:circ_axial_density}, known up to a vector of parameters $\bm{\theta}_j=(\bm{\gamma}_j, \bm{\alpha}_j, \rho_j)$, where the vector $\bm{\gamma}_j$ includes the parameters of the marginal circular distribution of the $j$th component, $\bm{\alpha}_j$ includes the parameters of the marginal axial distribution of the $j$th component, and, finally, $\rho_j$ indicates the correlation between the axial and the circular variables within the $j$-th component. 

Model \eqref{eq:mixture1} allows fitting heterogeneous samples by segmenting the data according to $J$ latent classes associated with different bivariate axial-circular densities. It however relies on a homogeneous set of class membership probabilities $\pi_1, \ldots, \pi_J$. This could be a shortcoming in our case study, as it is quite likely that latent classes are associated with the topography of the observation site, such as the cone aspect and slope steepness. A useful extension of \eqref{eq:mixture1} that integrates topographic information as desired is obtained by assuming that the conditional distribution of each observation $(x,y)$ given a vector $\bm{z}$ of $q$ topographic covariates  depends on nonhomogeneous class membership probabilities $\pi_j(\bm{z}; \bm{\beta})$, known up to a battery of multinomial regression parameters $\bm{\beta}=(\bm{\beta_2}, \ldots, \bm{\beta}_J)$, namely  
\begin{equation}\label{eq:mixture2}
f(x,y \mid \bm{z}) = \sum_{j=1}^{J}\pi_j(\bm{z}; \bm{\beta}) f(x,y; \bm{\theta}_j) 
\end{equation}
where
\[\log \frac{\pi_j(\bm{z})}{\pi_1(\bm{z})}=\bm{z}^{\sf T}\bm{\beta}_j, \qquad j= 2, \ldots, J.
\]

\section{Maximum likelihood estimation}\label{sec:estimation}
Let $\bm{x}=(x_1, \ldots, x_n)$ and  $\bm{y}=(y_1, \ldots, y_n)$ be two vectors of circular and axial data, respectively, each associated with a vector $\bm{z}_i$ of covariate information. Under our proposal, the likelihood function is given by 
\begin{equation}\label{eq:likelihood}
L(\bm{\beta}, \bm{\theta} ) = \prod_{i = 1}^n \sum_{j = 1}^J \pi_j(\bm{z}_i; \bm{\beta})f(x_i, y_i; \bm{\theta}_j).
\end{equation}
An Expectation-Maximization (EM) can be exploited to maximize \eqref{eq:likelihood}. It is based on the following complete-data log-likelihood function

\begin{equation}\label{eq:loglik_comp}
    \log L_c(\bm{\beta}, \bm{\theta}) = \sum_{i = 1}^n \sum_{j = 1}^J u_{ij}[\log \pi_j(\bm{z}_i; \bm{\beta}) + \log f(x_i, y_i; \bm{\theta}_j)],
\end{equation}
where $u_{ij}$ is the class membership of observation $i$ and it is equal to 1 if the observation belongs to class $j$ and 0 otherwise. 
The algorithm is iterated by alternating an E step and an M step. Given the estimates $\hat{\bm{\beta}}^{(s)}$ and $\hat{\bm{\theta}}^{(s)}$ obtained at the previous iteration, the E step evaluates the conditional expected value of the class membership given the observed data, say
\begin{equation}
\label{eq:E_step}
\hat{u}_{ij} = \frac{\pi_{j}(\bm{z}_i; \hat{\bm{\beta}}^{(s)})f(x_i, y_i; \hat{\bm{\theta}}_{j}^{(s)})}
{\sum_{j=1}^J \pi_j(\bm{z}_i; \hat{\bm{\beta}}^{(s)})f(x_i, y_i; \hat{\bm{\theta}}_{j}^{(s)})}, \, \quad i=1, \ldots n, \quad j = 1, \dots, J.
\end{equation}
The M step updates the parameter estimates by maximizing the expected value $Q(\bm{\beta},\bm{\theta})$ of the complete-data log-likelihood, obtained by plugging the estimated class memberships \eqref{eq:E_step} into \eqref{eq:loglik_comp}, say
\begin{align}\label{eq:M_step}
      Q(\bm{\beta},\bm{\theta}) =& \sum_{i = 1}^n \sum_{j = 1}^J \left[ \hat{u}_{ij}\log \pi_j(\bm{z}_i; \bm{\beta}) + \hat{u}_{ij}\log f(x_i, y_i; \bm{\theta}_j)\right] \nonumber\\
      =&\sum_{i = 1}^n \sum_{j = 1}^J \left[ \hat{u}_{ij}\log \pi_j(\bm{z}_i; \bm{\beta}) \right] +   \sum_{i = 1}^n \sum_{j = 1}^J \left[\hat{u}_{ij}\log f(x_i, y_i; \bm{\theta}_j)\right]\nonumber \\
      =&Q(\bm{\beta})+Q(\bm{\theta})
\end{align}
The function $Q(\bm{\beta},\bm{\theta})$ is the sum of two functions that depend on independent sets of parameters and can be therefore maximized independently. In particular, the first term of the sum, $Q(\bm{\beta})$, corresponds to a weighted multinomial log-likelihood that is maximized by fitting a weighted multinomial logistic regression. The second term of the sum, $Q(\bm{\theta})$, can be instead maximized using an inference function for margins approach \citep[IFM; ][]{Kim_copulas2007, Lagona_copulas2019}. By recalling \eqref{eq:parametric density}, $Q(\bm{\theta})$ can be written as the sum of three components, namely 
\begin{eqnarray}
\sum_{i=1}^{n}\hat{u}_{ij}\log f(x_i,y_i; \bm{\theta}_j)&=\sum_{i=1}^{n}&\hat{u}_{ij}\log c_{\rho_j}(F(x_i; \bm{\gamma}_j), F(y_i;\bm{\alpha}_j))\label{eq:rho}\\
&+&\sum_{i=1}^{n}\hat{u}_{ij}\log f(x_i; \bm{\gamma}_j)\label{eq:gamma}\\
&+&\sum_{i=1}^{n}\hat{u}_{ij}\log f(y_i; \bm{\alpha}_j)\label{eq:alpha}.
\end{eqnarray}
IFM proceeds accordingly by finding first the parameter values $\hat{\bm{\gamma}}$ and $\hat{\bm{\alpha}}$ that respectively maximize (\ref{eq:gamma}) and (\ref{eq:alpha}). Then, the function (\ref{eq:rho}) is evaluated at $\bm{\gamma}=\hat{\bm{\gamma}}$ and $\bm{\alpha}=\hat{\mathbf{\alpha}}$ and maximized with respect to $\rho_j$, $j=1, \ldots, J$. 

Uncertainty quantification for the parameter estimates is based on a parametric bootstrap approach. Specifically, we re-fitted the model to $B$ bootstrap samples, which were simulated from the proposed model by fixing the \textit{true} parameters to the MLE estimates. This yield $\lbrace \bm{\theta} \rbrace_{b=1}^B$ and $\lbrace \bm{\beta} \rbrace_{b=1}^B$ sets of estimates, exploited to compute equal-tail (ET) confidence intervals. Within the proposed framework, this inferential approach proves convenient as it overcomes the non-trivial computation of the observed information matrix. As an aside, we note that symmetric confidence intervals are not guaranteed to be within the admissible parameter space when an estimate is close to the space boundary. Equal-tail, bootstrap-based confidence intervals do not suffer from this drawback.

\section{Simulation study}
\label{sec:simulation}
We run a simulation study to check the ability of the proposed model to recover the population parameter values under different scenarios, as well as to assign each observation a probability of coming from one of the latent classes, and eventually classify it. Simulating from the model in \eqref{eq:mixture2} is straightforward and a detailed description of the algorithm is available in the Appendix (Algorithm \ref{algosim}). 

Precisely, we generate $N=200$ replicas, each with $n = 600$ observations for $J \in \lbrace 2,3\rbrace$, including 2 covariates. Parameters used for the simulation are available in the left columns of Table \ref{tab:simresdensity} and Table \ref{tab:simresbeta}. While the values for the density parameters have been chosen to explore the whole range of the data domain, the values of the covariates have been randomly generated from $N(0, 4)$ and $Ber(0.5)$ and the corresponding $\bm{\beta}$s from $N(0, 4)$. These scenarios -- though not exhaustive -- aim to mimic possible real data situations where mixture components may (or may not) overlap and show positive (or negative) correlations, while also accounting for different effects of covariates on the membership probabilities. Figure \ref{fig:simulatedex} of the Appendix displays two examples of simulated datasets under the considered scenarios for $J=2$ and $J=3$, respectively.

\begin{figure}[ht]
        \centering
        \includegraphics[width=0.6\linewidth]{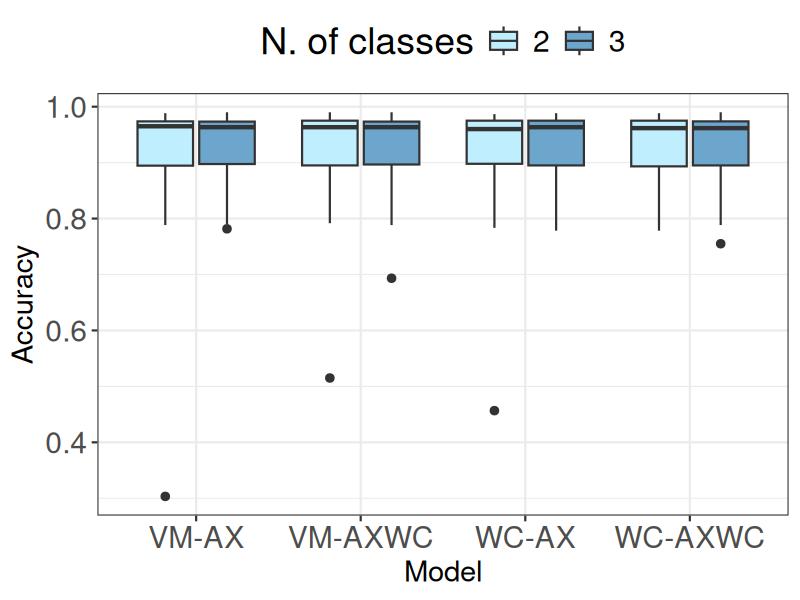}
        \caption{Accuracy distribution across the 200 replicas for each simulated scenario (VM = circular von Mises; AX = axial von Mises; WC = circular wrapped Cauchy; AXWC = axial wrapped Cauchy).}
        \label{fig:aris}
    \end{figure}

We report the mean and the equal-tail (ET) 95\% confidence interval for each parameter obtained by averaging across the replicas. 
This scheme is implemented for each axial-circular combination with marginals in Table \ref{tab:univariate_densities}. 
Results of the simulation study about the recovery of the density and regression parameters are reported in Table \ref{tab:simresdensity} and Table \ref{tab:simresbeta}, respectively. They show that in all cases the point estimates are accurate and that the true values are always included in the ET 95\% confidence intervals, highlighting the ability of the proposed model to recover the parameters for both the observed and the latent component. Results of the classification performances are instead illustrated in Figure \ref{fig:aris}. The accuracy, i.e., the percentage of points correctly allocated to the right latent class, is always very high (above 80\% in all cases with a few exceptions). 

\begin{sidewaystable}[ht]
    \centering
    \resizebox{.95\textwidth}{!}{%
    \begin{tabular}{lc|ccccccccccc}
    \toprule
     \multirow{2}{*}{Family} & \multirow{2}{*}{$j$} & \multicolumn{2}{c}{$\mu_{circ}$} & \multicolumn{2}{c}{$\kappa_{circ}$} &  \multicolumn{2}{c}{$\mu_{ax}$} & \multicolumn{2}{c}{$\kappa_{ax}$} & \multicolumn{2}{c}{$\rho$}\\
      \cline{3-12}
      &  & true & est & true & est & true & est & true & est & true & est \\
      \midrule
      \multirow{5}{*}{VM-AX} & 1 &  1 & 1.00  (0.92, 1.09)  & 2 & 2.03  (1.74,  2.31) & 0.5 & 0.50 (0.40, 0.61) & 2 & 2.03  (1.70,  2.34) & -0.45 & -0.45 (-0.50, -0.39) \\ 
        & 2  & 5 & 5.00 (4.94, 5.07) & 6 & 6.19 (4.87, 8.16) & 2 & 2.00 (1.93, 2.06) & 5 & 5.12 (4.06, 6.31) & 0.60 & 0.58 (0.49, 0.65)\\
        \cline{2-12}
       & 1 &  1 & 1.00  (0.90, 1.10)  & 3 & 3.09  (2.48,  3.87) & 0.5 & 0.49 (0.35, 0.62) & 2 & 2.08  (1.54,  2.62) & -0.45 & -0.39 (-0.50, -0.11)\\ 
       &  2  & 5 & 5.00 (4.92, 5.07) & 5 & 5.05 (3.69, 6.71) & 2 & 2.00 (1.92, 2.07) & 5 & 5.14 (3.96, 6.79) & 0.60 & 0.57 (0.46, 0.64)\\
       &  3  & 3 & 3.00 (2.95, 3.04) & 10 & 10.10 (8.31, 12.70) & 1.5 & 1.50 (1.45, 1.54) & 9 & 9.13 (7.63, 11.10) & 0.10 & 0.09 (-0.06, 0.17)\\
       \midrule
      \multirow{5}{*}{VM-AXWC} & 1 &  1 & 0.99  (0.86, 1.11)  & 3 & 3.03  (1.99,  3.92) & 0.5 & 0.61 (0.13, 2.86) & 0.3 & 0.31  (0.15,  0.43) & -0.45 & -0.43 (-0.52, -0.36)\\ 
       &  2  & 5 & 4.98 (4.91, 5.08) & 5 & 5.24 (3.60, 6.83) & 2 & 2.00 (1.83, 2.15) & 0.55 & 0.56 (0.45, 0.65) & 0.60 & 0.57 (0.45, 0.66)\\
         \cline{2-12}
       & 1 &  1 & 1.00  (0.92, 1.10)  & 2 & 2.04  (1.68,  2.40) & 0.5 & 0.51 (0.20, 0.83) & 0.3 & 0.31  (0.13,  0.41) & -0.45 & -0.41 (-0.49, -0.31)\\ 
       &  2  & 5 & 4.98 (4.93, 5.07) & 6 & 5.99 (4.60, 7.37) & 2 & 2.00 (1.92, 2.08) & 0.7 & 0.70 (0.64, 0.75) & 0.60 & 0.58 (0.50, 0.64)\\
       &  3  & 3 & 3.00 (2.95, 3.04) & 10 & 10.02 (8.31, 12.30) & 1.5 & 1.50 (1.48, 1.52) & 0.9 & 0.90 (0.88, 0.92) & 0.10 & 0.09 (-0.08, 0.18)\\
        \midrule
      \multirow{5}{*}{WC-AX} &  1 &  1 & 1.00 (0.76, 1.24)  & 0.3 & 0.30 (0.25, 0.37) & 0.5 & 0.50 (0.37, 0.61) & 2 & 2.02 (1.69, 2.42) & -0.45 & -0.44 (-0.50, -0.38)\\ 
      &   2  & 5 & 5.00 (4.98, 5.02) & 0.9 & 0.90 (0.88, 0.92) & 2 & 2.00  (1.95, 2.05) & 5 & 5.07 (4.37, 6.00) & 0.60 & 0.59 (0.53, 0.64)\\
       \cline{2-12}
      &  1 &  1 & 1.00 (0.71, 1.29)  & 0.3 & 0.31 (0.21, 0.40) & 0.5 & 0.50 (0.33, 0.67) & 2 & 2.11 (1.55, 2.81) & -0.45 & -0.43 (-0.52, -0.34)\\ 
      &   2  & 5 & 5.00 (4.98, 5.03) & 0.9 & 0.90 (0.87, 0.92) & 2 & 2.00  (1.92, 2.07) & 5 & 5.21 (3.65, 7.18) & 0.60 & 0.58 (0.46, 0.66)\\
      &   3  & 3 & 3.00 (2.86, 3.16) & 0.5 & 0.50 (0.41, 0.58) & 1.5 & 1.50  (1.45, 1.55) & 9 & 9.19 (7.44, 11.70) & 0.10 & 0.09 (-0.06, 0.20)\\
      \midrule
      \multirow{5}{*}{WC-AXWC} &  1 &  1 & 1.01 (0.82, 1.23)  & 0.3 & 0.31 (0.23, 0.37) & 0.5 & 0.49 (0.20, 0.80) & 0.3 & 0.31 (0.16, 0.41) & -0.45 & -0.42 (-0.50, -0.29)\\ 
      &   2  & 5 & 5.00 (4.98, 5.02) & 0.9 & 0.90 (0.88, 0.92) & 2 & 2.00  (1.92, 2.06) & 0.7 & 0.71 (0.64, 0.76) & 0.60 & 0.58 (0.51, 0.65)\\
      \cline{2-12}
      &  1 &  1 & 1.05 (0.70, 1.43)  & 0.3 & 0.31 (0.18, 0.42) & 0.5 & 0.55 (0.13, 1.21) & 0.3 & 0.33 (0.14, 0.51) & -0.45 & -0.40 (-0.51, -0.22)\\ 
      &   2  & 5 & 5.00 (4.98, 5.03) & 0.9 & 0.90 (0.87, 0.92) & 2 & 2.00  (1.81, 2.13) & 0.55 & 0.57 (0.46, 0.66) & 0.60 & 0.57 (0.46, 0.66)\\
      &   3  & 3 & 3.01 (2.87, 3.14) & 0.5 & 0.50 (0.39, 0.59) & 1.5 & 1.49  (1.48, 1.52) & 0.9 & 0.90 (0.87, 0.92) & 0.10 & 0.09 (-0.05, 0.19)\\
      \bottomrule
    \end{tabular}
    }
    \caption{Results of the simulation: recovery of the density parameters (VM = circular von Mises; AX = axial von Mises; WC = circular wrapped Cauchy; AXWC = axial wrapped Cauchy)}
    \label{tab:simresdensity}
\end{sidewaystable}

\begin{table}[ht]
    \centering
    \begin{tabular}{lc|cccccc}
    \toprule
       Family & $j$ & \multicolumn{2}{c}{$\beta_0$} & \multicolumn{2}{c}{$\beta_1$} & \multicolumn{2}{c}{$\beta_2$} \\ \cline{3-8}
       & & true & est & true & est & true & est \\
        \midrule
       VM-AX &  2 & -2.41 & -2.43 (-2.94, -2.00) & 0.55 & 0.56 (0.43, 0.71) & 2.17 & 2.18 (1.72, 2.77) \\
         \midrule
        & 2 & -0.09 & -0.06 (-0.51, 0.38) & 0.64 & 0.66 (0.50, 0.83) & 0.12 &  0.10 (-0.43,  0.60)\\
        & 3 & 1.32 & 1.36 (1.01, 1.75) & 1.17 & 1.19 (1.01, 1.46)  & -2.93 & -2.96 (-3.67, -2.37)  \\
        \midrule
        VM-AXVC  & 2 & -0.86 & -0.86 (-1.11, -0.55) & 0.23 & 0.23 (0.14, 0.33) & 0.37 &  0.37 (0.01,  0.71)\\
        \midrule
        & 2 & -0.86 & -0.88 (-1.31, -0.55) & 0.37 & 0.37 (0.26, 0.49) & 0.54 &  0.55 (-0.02,  1.13)\\
        & 3 & 0.23 & 0.22 (-0.10, 0.45) & -0.07 & -0.08 (-0.17, 0.01)  & -0.19 & -0.20 (-0.60, 0.17)  \\
        \midrule
        WC-AX & 2 & -1.26 & -1.32 (-2.10, -0.67) & 3.67 & 3.87 (2.82, 5.17) & 0.69 & 0.75 (-0.10, 1.72) \\
         \midrule
       &  2 & -0.09 & -0.07 (-0.66, 0.59) & 0.64 & 0.65  (0.45, 0.91) & 0.12 & 0.11 (-0.62, 0.79)\\
       &  3 & 1.32 & 1.36 (0.87, 1.91) & 1.17 & 1.21 (0.95, 1.56)  & -2.93 & -3.02 (-3.93, -2.13) \\
       \midrule
       WC-AXWC & 2 & -0.86 & -0.87 (-1.40, -0.38) & 0.23 & 0.25  (-0.14, 0.77) & 0.37 & 0.38  (-0.06, 0.79) \\
         \midrule
       &  2 &  -0.86 & -0.87 (-1.18, -0.552) & 0.37 & 0.38  (0.21, 0.55) & 0.54 & 0.55  (-0.03, 1.21) \\
       &  3 & 0.23 & 0.25 (-0.14, 0.77) & -0.07 & -0.08 (-0.22, 0.06)  & -0.19 & -0.23 (-0.77, 0.32) \\
         \bottomrule         
    \end{tabular}
    \caption{Results of the simulation study: recovery of the regression parameters (VM = circular von Mises; AX = axial von Mises; WC = circular wrapped Cauchy; AXWC = axial wrapped Cauchy )}
    \label{tab:simresbeta}
\end{table}

\section{Results}
\label{sec:results}
The mixture model \eqref{eq:mixture2} has been estimated on the real data described in Section \ref{sec:data}, by considering the axial direction of vegetation stripe and the circular direction of wind at $n=133$ locations on Marion Island. We include the aspect (North-East, North-West, South-East, South-West) and the slope as topographic covariates to model the class membership probabilities. The former identifies the side (NE, NW, SE or SW) of the scoria cone where each data point is located, while the latter measures the inclination (i.e. slope steepness) of the terrain surface (from 0 to 90 degrees). Our proposal allows to choose any pair of axial and circular marginal densities and, therefore, we estimated the model by considering four different specifications that combine either a circular von Mises or a circular wrapped Cauchy with either an axial von Mises or an axial wrapped Cauchy. These four model specifications have been estimated by increasing the number of mixture components $J$ from 2 to 4, therefore obtaining a battery of 12 estimated models. We select the best model as the one that minimizes the BIC. Results are reported in Table \ref{tab:modelsel}, which highlights that the model with $J=3$ components and the von Mises specification for both the circular and axial part provides the best compromise between goodness of fit and parsimony.

\begin{figure}[ht]
    \centering
    \begin{subfigure}[b]{.45\textwidth}
    \includegraphics[width = .99\textwidth]{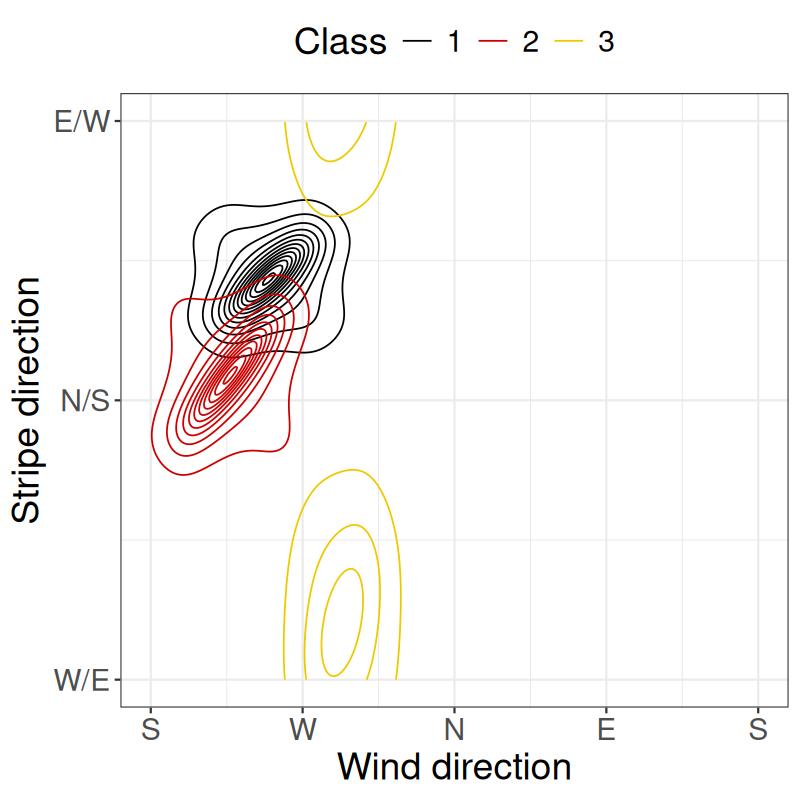}
    \caption{}
    \label{contourmod}
    \end{subfigure}
    \begin{subfigure}[b]{.45\textwidth}
    \includegraphics[width = .99\textwidth]{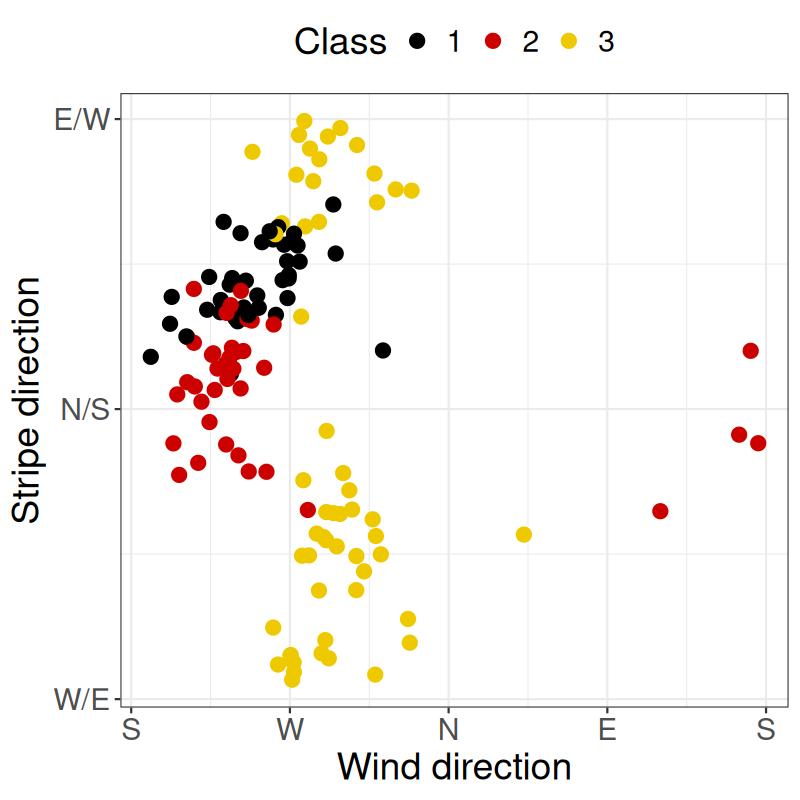}
    \caption{}
    \label{classmod}
    \end{subfigure}
    \caption{(a) Estimated contours and (b) resulting classification for the model with $J=3$.}
\end{figure}

\begin{figure}[ht]
        \centering
        \includegraphics[width=0.95\linewidth]{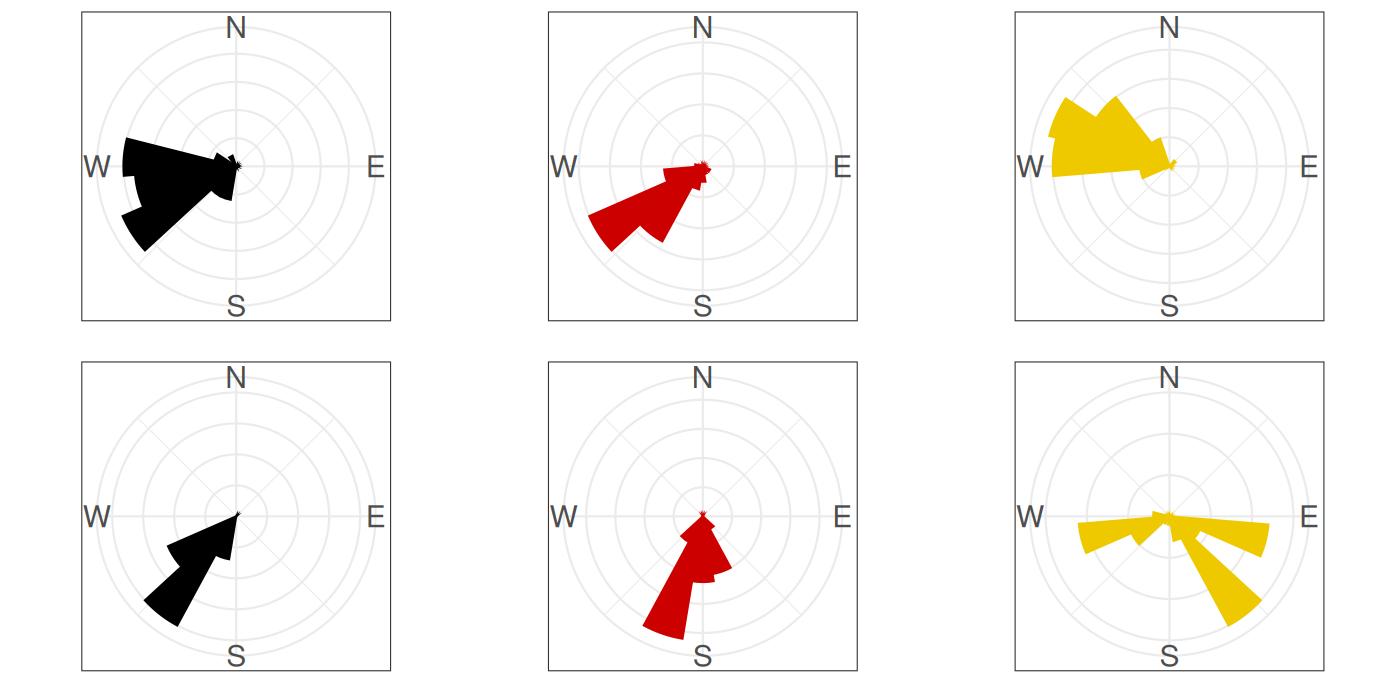}
        \caption{\label{fig:univ_classificationbest} Conditional univariate distribution of wind (top panels) and stripe (lower panels) directions, given the allocation within three latent classes.}
\end{figure}

\begin{figure}[ht]
    \centering
    \begin{subfigure}[b]{.45\textwidth}
    \includegraphics[width = .99\textwidth]{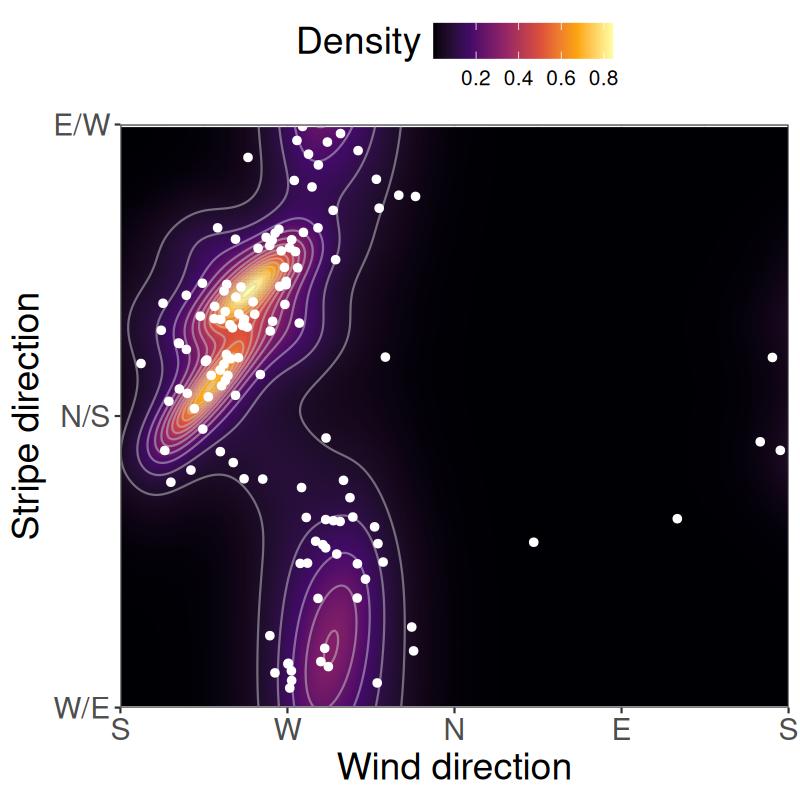}
    \caption{}
    \label{jointmod}
    \end{subfigure}
    \begin{subfigure}[b]{.45\textwidth}
    \includegraphics[width = .99\textwidth]{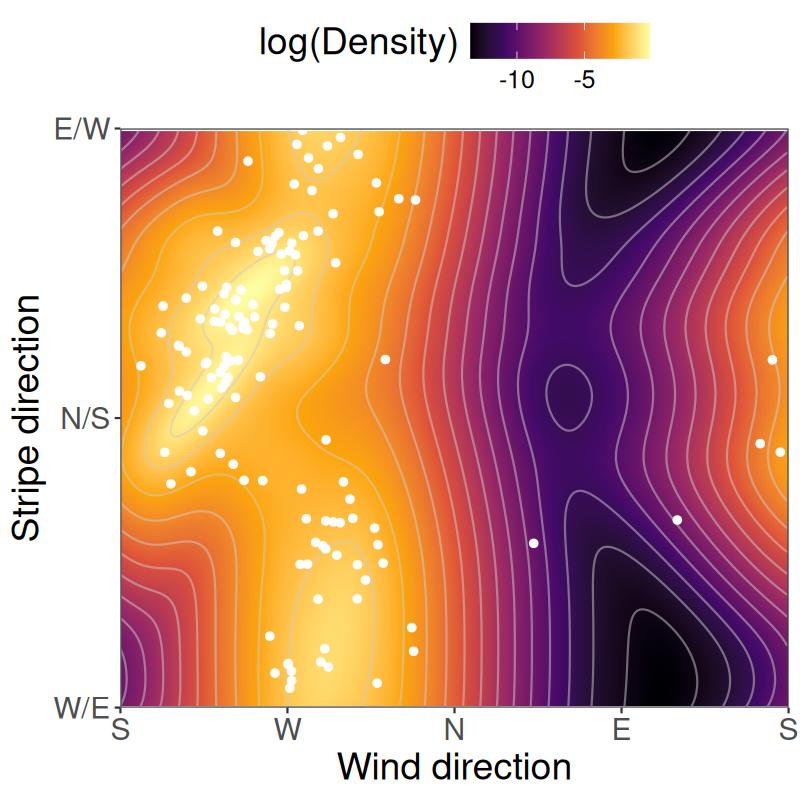}
    \caption{}
    \label{logjointmod}
    \end{subfigure}
     \caption{Estimated joint density (a) and on the log-scale (b), as estimated by a mixture of $J=3$ axial-circular densities.}
    \label{fig:classificationbest}
    \end{figure}

Parameters estimates of the best model are reported in Table \ref{tab:parestbest} and Table \ref{tab:parestbeta}, together with the ET confidence intervals obtained using $B = 1000$ bootstrap samples. Figure \ref{contourmod} and Figure \ref{classmod} display the contours of the three components and the resulting segmentation of the data, obtained by exploiting the class memberships probabilities in Eq. \eqref{eq:E_step} at the last EM iteration and allocating each data point to the class with the maximum probability. Figure \ref{fig:univ_classificationbest} displays the univariate distributions of the data, conditional on such allocation. By computing the weights $p_j=\sum_{i=1}^{n}\hat{u}_{ij}/n$, we obtain the estimated bivariate distributions $\sum_{j}p_jf(x,y; \hat{\bm{\theta}}_j)$. Figures \ref{jointmod} and \ref{logjointmod} show the levels-plot of the distribution in the real and the log-scale respectively, overlapped on the observed data, indicating a reasonable goodness of fit, which can be improved by choosing a larger number of components if desired.

 \begin{table}[ht]
    \centering
    \begin{tabular}{lr|ccc|ccc}
    \toprule
        Model & & \multicolumn{3}{c}{logL} &  \multicolumn{3}{c}{BIC}  \\
         \cline{3-8}
        & J & 2 & 3 & 4 & 2 & 3 & 4 \\
       \midrule
       VM-AX & &-182.14 & \textbf{-156.02} & -147.32 & 427.86 & \textbf{419.62 }& 446.23 \\
       WC-AX & &-185.50 & -162.28 & -139.66  & 434.58 &  432.15 & 430.93\\
       VM-AXWC && -192.88 & -167.05 & -149.59 & 460.53 &  454.43 & 468.56 \\
       WC-AXWC & &-192.87 & -167.83 & -148.57 & 460.52  & 456.00 & 470.65 \\
       \bottomrule
    \end{tabular}
    
    \caption{Log-likelihood and BIC for each estimated model (VM = circular von Mises; AX = axial von Mises; WC = circular wrapped Cauchy; AXWC = axial wrapped Cauchy)}
    \label{tab:modelsel}
\end{table}

 \begin{table}[ht]
    \centering
    \resizebox{.95\textwidth}{!}{%
    \begin{tabular}{l|ccccc}
    \toprule
       $j$  &  $\mu_{{\rm circ}}$ &  $\kappa_{{\rm circ}}$ &  $\mu_{{\rm axial}}$ &  $\kappa_{{\rm axial}}$ & $\rho$\\
       \midrule
        1 & -1.92 (-2.05, -1.80) & 5.60 (4.59, 10.06) & 0.68 (0.62, 0.75) & 18.96 (14.8, 36.8) & 0.24 (0.09, 0.42) \\
        2 & -2.32 (-2.45, -2.20) & 5.00 (4.26, 9.02) & 0.14 (0.05, 0.23) & 10.01 (8.44, 19.2) & 0.35 (0.12, 0.50) \\
        3 & -1.16 (-1.25, -1.05) & 5.99 (5.33, 10.20) & -1.25 (-1.39, -1.09) & 3.13 (2.60, 5.30) & 0.07 (-0.16, 0.22) \\
        \bottomrule
    \end{tabular}
    }
    \caption{Point estimates (ET confidence intervals) for the parameters of the density.}
    \label{tab:parestbest}
\end{table}
\begin{table}[ht]
    \centering
    \resizebox{.95\textwidth}{!}{%
    \begin{tabular}{l|ccccc}
    \toprule
    $j$ & NE & $\Delta$NW & $\Delta$SE & $\Delta$SW & Slope \\
    \midrule
    2 &  -15.30 (-269.00, -4.63) & -6.81 (-43.2, 6.37) & 11.40 (5.16, 164.00) & -11.01 (-172.00, -3.57) & 0.30 (0.03, 8.81) \\
   3 &  7.35 (0.76, 67.00) & -1.50 (-9.28, 45.10) & -17.12 (-104.00, -13.4) & -10.16 (-71.10, -4.94) & 0.05 (-0.03, 0.23) \\
     \bottomrule
    \end{tabular}
    }
    \caption{Estimated regression coefficients. Reference class is $j = 1$.}
    \label{tab:parestbeta}
\end{table}

 Within two latent classes ($j=1,2$), we obtain strong positive correlations between dominant wind direction and vegetation stripe orientation. The first component, comprising locations chiefly located on southwestern aspects of scoria cones, reveals a positive correlation ($\approx$ 0.25), with vegetation stripes orientated towards the northeast and winds predominantly blowing from the southwest. The second component, comprising locations on the southeastern aspects of scoria cones, also exhibits a positive correlation ($\approx 0.35$), although for this component the vegetation stripes primarily have a north-south orientation. Thus, this second component behaves similarly to the first, except that the stripes are orientated to dominant winds which have a more southerly direction. We also observe that scoria cone surfaces are steeper for the second component, reflecting significantly greater slope angles for the southeastern aspects. 

In contrast, the third class, comprising locations chiefly from the northern aspects of scoria cones, does not show a significant correlation between wind direction and stripe orientation. On the northeastern aspect of scoria cones, vegetation stripes were orientated along both northeast and the northwest axes, despite wind direction being fairly consistently from the northwest. A similar pattern, although with more variation, was observed for vegetation stripes on the northwestern aspects of scoria cones; vegetation stripes' orientation spanned across almost all axes (with only north-south orientated stripes being rare). Therefore, while wind direction was relatively consistent on this aspect of scoria cones (westerly to north-north-westerly winds), stripe direction was remarkably variable. For this component, the effect of slope was negligible.

Our findings, based on data collected from scoria cones occurring across a range of environmental conditions on Marion Island, suggest that wind direction likely contributes strongly to determining vegetation stripe orientation, although only on the southern aspects of scoria cones. Dominant winds likely shape the orientation of vegetation stripes through a mechanism of neighbouring plants providing wind shelter to downwind individuals. Given the wide range of negative physiological and morphological impacts that strong winds can have on plants (e.g. abrasion of leaf surfaces and the resulting increase in water loss, slowing of photosynthetic rates, mechanical breakage; reviewed in \citet{gardiner2016wind,de2008effects}), any form of wind shelter could potentially strongly impact plant growth and survival in this system \citep[see, e.g., how wind impacts plant communities in this system;][]{momberg2021species}. As a result, where plants are able to ameliorate wind stress, downwind (i.e. sheltered) plants likely experience improved performance, leading to a directional establishment of vegetation in the lee of existing vegetation \citep[as suggested for linear "ribbon" forests;][]{momberg2021exposing} or any features that block dominant winds \citep[e.g. ][]{bekker2008linear}.

The lack of a clear correlation between wind direction and vegetation stripe orientation on northern aspects of scoria cones was unexpected and suggests some key environmental differences between the northern and southern aspects. One potential explanation for this difference could be that higher wind speeds are experienced on the northern aspects (as expected, e.g., on the north-western aspect of the island), and that the mechanisms that potentially give rise to stripe formation (e.g. where neighbouring individuals provide some degree of wind sheltering to reduce the stress and disturbance of adjacent plants) are overwhelmed by stronger winds (i.e. the concept of collapse of facilitation; \cite{michalet2014two}). In other words, on northern aspects, wind speeds are possibly so stressful that neighbouring individuals are unable to provide enough protection from winds that vegetation stripe orientation is driven by other mechanisms. However, CFD simulations suggest that wind speed differs more between eastern (weaker) and western (stronger) aspects, than between northern and southern aspects.

An alternative explanation for the contrasting results from northern and southern scoria cone aspects could be that differences in incident solar radiation \citep{mccune2002equations} and, as a result, surface temperatures, cause plants to have differential vulnerability to the dominant winds. As equator-facing slopes receive more direct sunlight than polar-facing slopes, they tend to be warmer and, therefore generally in cold environments, provide more favourable conditions for plant growth and survival. Indeed, on sub-Antarctic Macquarie Island the cover of several plant species has been observed to vary with aspect \citep{kirkpatrick2002change, le2008rapid}, possibly reflecting different thermal preferences (although this likely also reflects differences in wind exposure). As a result, the lack of a strong relationship between vegetation stripe orientation and wind direction on northern scoria cone aspects could reflect warmer temperatures creating a more favourable environment, and reducing the benefits from (or need for) sheltering by neighbouring plants.   

In contrast to wind direction and aspect, our results demonstrate a weak influence of slope angle on vegetation stripes, in agreement with, for example, \cite{deblauwe2008global} who examined the occurrence of vegetation stripes in semi-arid areas. Therefore, despite the potential for slope angle to alter the rate of substrate and/or water movement, this does not appear to be a strong enough environmental influence to control vegetation stripe orientation. However, the influence of slope steepness may also potentially be related to the relative orientation of wind direction and slope direction. For example, if the dominant wind is blowing directly up a slope (i.e. parallel with the slope aspect), we would expect vegetation stripes to be orientated vertically (i.e. a strong positive correlation), irrespective of slope steepness. However, in the opposite scenario where the wind is blowing across slope (i.e. perpendicular to the slope aspect), we would anticipate vegetation stripes that follow the same orientation as the wind chiefly on flatter surfaces. Instead, on steeper slopes, the orientation of the vegetation stripes should reflect the influence of both wind direction and the movement of the substrate downslope. Therefore, while the slope steepness is clearly not a strong driver of vegetation stripe orientation in this system, there are still potential mechanisms of influence that could be examined.

\section{Discussion}
\label{sec:discussion}
Vegetation stripes are a widely distributed ecological occurrence, typically observed in wind, cold and/or dry environments. Due to the potential for vegetation stripes to affect ecosystem functioning in climatically extreme environments, for example, by altering the movement of sediment and water and by influencing plant survival and productivity \citep{ludwig1999stripes, morgan2010wind}, there has been considerable interest in the mechanisms contributing to their formation and maintenance. Indeed, recent suggestions that patterned vegetation (including vegetation stripes, but also forms of patchy vegetation) may be particularly sensitive to changing environmental conditions (including climate change), and therefore can potentially act as early warning signals for environmental tipping points and ecosystem collapse \citep{rietkerk2021evasion}.

Several linear environmental features have been reported from the sub-Antarctic, including sorted stone stripes (i.e. an unvegetated feature), vegetated terraces, and vegetation stripes. While the orientation of sorted stripes has been linked to wind direction, and there has been much speculation about the influence of wind on vegetated terraces \citep[i.e. stepped features in the landscape with vegetation growing on the flat surface; see e.g. ][]{loffler1984macquarie, selkirk1998active}, the joint distribution of wind direction and vegetation stripe orientation in the sub-Antarctic has not been explored prior to this research. 

The goal of modelling the joint distribution of wind direction and vegetation orientation is complicated by the nonstandard support of the data (the cartesian product of a circle and a semi-circle) and has been here accomplished by a novel class of bivariate distributions with mixed axial and circular margins. While the literature offers several examples of univariate circular and axial distributions, this is, to the best authors' knowledge, the first proposal of a bivariate axial-circular distribution. Our distribution fulfills the periodicity constrains required by the mixed axial-circular setting; it flexibly accommodates any marginal axial and circular marginal distributions; it allows for interpretable and computationally inexpensive inference. 

If required, the simple structure of the proposed distribution makes it an excellent candidate as a building block of more general models. In our cases study, for example, the heavy heterogeneity of the data suggested to deploy the proposed distribution as the generic component of a mixture model, with nonhomogeneous class-membership probabilities. Under this setting, the model was capable to segment mixed axial and circular data according to intuitively appealing latent classes, providing a parsimonious description of vegetation patterns in terms of interpretable environmental regimes.




\subsection*{Financial disclosure}
This work has been partially supported by MIUR, grant number 2022XRHT8R - The SMILE project: Statistical Modelling and Inference to Live the Environment, by the National Research Foundation (NRF) of South Africa, Reference: SRUG2204203965, and by Subcommittee A of the Research Committee from the Faculty of Economic and Management Sciences at Stellenbosch University.

The collection of the data analysed here was supported by the National Research Foundation's South African National Antarctic Programme (grant numbers 93077, 110726 and 110723) and was conducted under permits from the Prince Edward Islands Management Committee (PEIMC1/2013).

\subsection*{Conflict of interest}

The authors declare no potential conflict of interest.

\section*{Supporting information}
All data and codes that have been used to obtain the results presented in this work are available at \url{https://github.com/minmar94/AxialMixtureRegression/}


\newpage
\section*{Appendix}

\begin{figure}[ht]
    \centering
    \begin{subfigure}[b]{.48\textwidth}
    \includegraphics[width = .99\textwidth]{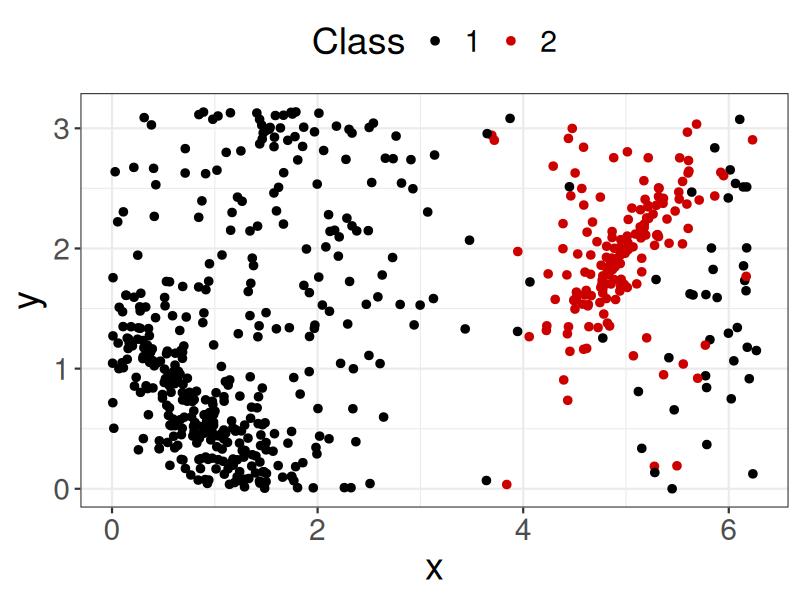}
    \caption{}
    \end{subfigure}
    \begin{subfigure}[b]{.48\textwidth}
    \includegraphics[width = .99\textwidth]{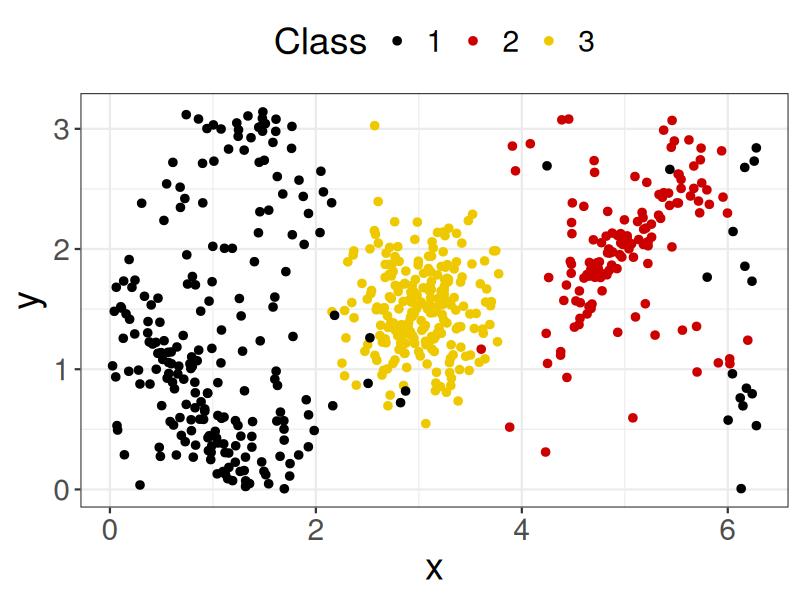}
    \caption{}
    \end{subfigure}
   \caption{Example of simulated dataset with (a) $J=2$ and (b) $J=3$ classes.}
   \label{fig:simulatedex}
    \end{figure}

\begin{algorithm}[ht]
\caption{Simulation from a mixture of copula-based axial-circular distributions}\label{algosim} 
\begin{algorithmic}
\Require a vector of covariates $\bm{z}$ and parameters $(\bm{\beta}, \bm{\gamma}, \bm{\alpha})$
\Ensure $n\geq 2$
\For{$i=1,\dots,n$}

\State \textbf{1: Generate predictor values}
\State \quad \textbf{a:} Compute $\eta_{ij} = \bm{z}_i^{\top}\bm{\beta}_j, \, $  $j = 2, \dots, J$
\State \quad \textbf{b:} Compute $\pi_{ij} = \frac{\exp(\eta_{ij})}{1+\sum_{j=2}^J \exp(\eta_{ij})}, \, $ $j = 2, \dots, J$
\State \quad \textbf{c:} Compute $\pi_{i1} = \frac{1}{1+\sum_{j=2}^J \exp(\eta_{ij})}$
\State \textbf{2: Simulate cluster membership} $u_{ij} \sim Multinom(\pi_{i1}, \dots, \pi_{iJ})$
\State \textbf{3: Simulate axial-circular observations}
\State \quad \textbf{a:} Draw $\nu \sim Unif(0,1)$
\State \quad \textbf{b:} Compute $y_i = F^{-1}_{\text{axial}}(\nu; \bm{\alpha_j})$
\State \quad \textbf{c:} Compute $\delta_1 = 2\pi F_{\text{axial}}(y_i; \bm{\alpha_j})$
\State \quad \textbf{d:} Draw $\delta_2 \sim f(\delta \mid \delta_1)$, \State \quad where $f(\delta \mid \delta_1)$ is the conditional WC obtained by \citet{Kato_Pewsey2015}, Th. 1, Sec. 2.4, with $\kappa_1=\kappa_2=0$
\State \quad \textbf{e:} Compute $x_i = F^{-1}_{\text{circ}}(2\pi^{-1} \delta_2; \bm{\gamma}_j)$
\EndFor
\end{algorithmic}
\end{algorithm}

\newpage

\begin{thebibliography}{46}
\providecommand{\natexlab}[1]{#1}
\providecommand{\url}[1]{\texttt{#1}}
\expandafter\ifx\csname urlstyle\endcsname\relax
  \providecommand{\doi}[1]{doi: #1}\else
  \providecommand{\doi}{doi: \begingroup \urlstyle{rm}\Url}\fi

\bibitem[Abuzaid(2018)]{abuzaid2018half}
A.~H. Abuzaid.
\newblock A half circular distribution for modeling the posterior corneal curvature.
\newblock \emph{Communications in Statistics-Theory and Methods}, 47\penalty0 (13):\penalty0 3118--3124, 2018.

\bibitem[Alftine and Malanson(2004)]{alftine2004directional}
K.~J. Alftine and G.~P. Malanson.
\newblock Directional positive feedback and pattern at an alpine tree line.
\newblock \emph{Journal of Vegetation Science}, 15\penalty0 (1):\penalty0 3--12, 2004.

\bibitem[Arnold and SenGupta(2006)]{Arnold_Sengupta2006}
B.~C. Arnold and A.~SenGupta.
\newblock Probability distributions and statistical inference for axial data.
\newblock \emph{Environmental and Ecological Statistics}, 13:\penalty0 271--285, 2006.
\newblock \doi{10.1007/s10651-004-0011-8}.

\bibitem[Baartman et~al.(2018)Baartman, Temme, and Saco]{baartman2018effect}
J.~E. Baartman, A.~J. Temme, and P.~M. Saco.
\newblock The effect of landform variation on vegetation patterning and related sediment dynamics.
\newblock \emph{Earth Surface Processes and Landforms}, 43\penalty0 (10):\penalty0 2121--2135, 2018.

\bibitem[Bekker and Malanson(2008)]{bekker2008linear}
M.~F. Bekker and G.~P. Malanson.
\newblock Linear forest patterns in subalpine environments.
\newblock \emph{Progress in Physical Geography}, 32\penalty0 (6):\penalty0 635--653, 2008.

\bibitem[Chown and Froneman(2008)]{chown2008prince}
S.~Chown and P.~W. Froneman.
\newblock \emph{The Prince Edward Islands: land-sea interactions in a changing ecosystem}.
\newblock African Sun Media, 2008.

\bibitem[Dayton and Macready(1988)]{dayton1988_concvariable_mixture}
C.~M. Dayton and G.~B. Macready.
\newblock Concomitant-variable latent-class models.
\newblock \emph{Journal of the American Statistical Association}, 83\penalty0 (401):\penalty0 173--178, 1988.
\newblock \doi{10.1080/01621459.1988.10478584}.
\newblock URL \url{https://www.tandfonline.com/doi/abs/10.1080/01621459.1988.10478584}.

\bibitem[De~Langre(2008)]{de2008effects}
E.~De~Langre.
\newblock Effects of wind on plants.
\newblock \emph{Annu. Rev. Fluid Mech.}, 40\penalty0 (1):\penalty0 141--168, 2008.

\bibitem[Deblauwe et~al.(2008)Deblauwe, Barbier, Couteron, Lejeune, and Bogaert]{deblauwe2008global}
V.~Deblauwe, N.~Barbier, P.~Couteron, O.~Lejeune, and J.~Bogaert.
\newblock The global biogeography of semi-arid periodic vegetation patterns.
\newblock \emph{Global Ecology and Biogeography}, 17\penalty0 (6):\penalty0 715--723, 2008.

\bibitem[García-Portugués et~al.(2013)García-Portugués, Crujeiras, and González-Manteiga]{garcia_portugues2013}
E.~García-Portugués, R.~M. Crujeiras, and W.~González-Manteiga.
\newblock Exploring wind direction and so2 concentration by circular- linear density estimation.
\newblock \emph{Stochastic Environmental Research and Risk Assessment}, pages 1055--1067, 2013.

\bibitem[Gardiner et~al.(2016)Gardiner, Berry, and Moulia]{gardiner2016wind}
B.~Gardiner, P.~Berry, and B.~Moulia.
\newblock Wind impacts on plant growth, mechanics and damage.
\newblock \emph{Plant science}, 245:\penalty0 94--118, 2016.

\bibitem[Goddard et~al.(2022)Goddard, Craig, Schoombie, and Le~Roux]{goddard2022investigation}
K.~Goddard, K.~Craig, J.~Schoombie, and P.~Le~Roux.
\newblock Investigation of ecologically relevant wind patterns on marion island using computational fluid dynamics and measured data.
\newblock \emph{Ecological Modelling}, 464:\penalty0 109827, 2022.

\bibitem[Hall(1979)]{hall1979sorted}
K.~Hall.
\newblock Sorted stripes orientated by wind action: some observations from sub-antarctic marion island.
\newblock \emph{Earth Surface Processes}, 4\penalty0 (3):\penalty0 281--289, 1979.

\bibitem[Hall(1994)]{hall1994some}
K.~Hall.
\newblock Some observations regarding sorted stripes, livingston island, south shetlands.
\newblock \emph{Permafrost and Periglacial Processes}, 5\penalty0 (2):\penalty0 119--126, 1994.

\bibitem[Haussmann et~al.(2009)Haussmann, McGeoch, and Boelhouwers]{haussmann2009interactions}
N.~Haussmann, M.~A. McGeoch, and J.~Boelhouwers.
\newblock Interactions between a cushion plant (azorella selago) and surface sediment transport on sub-antarctic marion island.
\newblock \emph{Geomorphology}, 107\penalty0 (3-4):\penalty0 139--148, 2009.

\bibitem[Hedding et~al.(2015)Hedding, Nel, and Anderson]{hedding2015aeolian}
D.~W. Hedding, W.~Nel, and R.~L. Anderson.
\newblock Aeolian processes and landforms in the sub-antarctic: preliminary observations from marion island.
\newblock \emph{Polar Research}, 34\penalty0 (1):\penalty0 26365, 2015.

\bibitem[Hodel and Fieberg(2022)]{hodel2022_cilindrical_copulas}
F.~H. Hodel and J.~R. Fieberg.
\newblock Circular--linear copulae for animal movement data.
\newblock \emph{Methods in Ecology and Evolution}, 13\penalty0 (5):\penalty0 1001--1013, 2022.

\bibitem[Holness(2001)]{holness2001orientation}
S.~D. Holness.
\newblock The orientation of sorted stripes in the maritime subantarctic, marion island.
\newblock \emph{Earth Surface Processes and Landforms: The Journal of the British Geomorphological Research Group}, 26\penalty0 (1):\penalty0 77--89, 2001.

\bibitem[Iftikhar et~al.(2022)Iftikhar, Ali, and Hanif]{iftikhar2022half}
A.~Iftikhar, A.~Ali, and M.~Hanif.
\newblock Half circular modified burr- iii distribution, application with different estimation methods.
\newblock \emph{Plos one}, 17\penalty0 (5):\penalty0 e0261901, 2022.

\bibitem[Imoto and Abe(2021)]{imoto_abe2021_copula}
T.~Imoto and T.~Abe.
\newblock Simple construction of a toroidal distribution from independent circular distributions.
\newblock \emph{Journal of Multivariate Analysis}, 186:\penalty0 104799, 2021.
\newblock ISSN 0047-259X.
\newblock \doi{https://doi.org/10.1016/j.jmva.2021.104799}.
\newblock URL \url{https://www.sciencedirect.com/science/article/pii/S0047259X21000774}.

\bibitem[Jones et~al.(2015)Jones, Pewsey, and Kato]{Jones_etal2015}
M.~C. Jones, A.~Pewsey, and S.~Kato.
\newblock On a class of circulas: copulas for circular distributions.
\newblock \emph{Annals of the Institute of Statistical Mathematics}, 67(5):\penalty0 843--862, 2015.

\bibitem[Kato and Pewsey(2015)]{Kato_Pewsey2015}
S.~Kato and A.~Pewsey.
\newblock {A Möbius transformation-induced distribution on the torus}.
\newblock \emph{Biometrika}, 102\penalty0 (2):\penalty0 359--370, 03 2015.
\newblock ISSN 0006-3444.
\newblock \doi{10.1093/biomet/asv003}.
\newblock URL \url{https://doi.org/10.1093/biomet/asv003}.

\bibitem[Kim et~al.(2007)Kim, Silvapulle, and Silvapulle]{Kim_copulas2007}
G.~Kim, M.~J. Silvapulle, and P.~Silvapulle.
\newblock Comparison of semiparametric and parametric methods for estimating copulas.
\newblock \emph{Computational Statistics \& Data Analysis}, 51\penalty0 (6):\penalty0 2836--2850, 2007.
\newblock ISSN 0167-9473.
\newblock \doi{https://doi.org/10.1016/j.csda.2006.10.009}.
\newblock URL \url{https://www.sciencedirect.com/science/article/pii/S0167947306003690}.

\bibitem[Kirkpatrick and Scott(2002)]{kirkpatrick2002change}
J.~Kirkpatrick and J.~Scott.
\newblock Change in undisturbed vegetation on the coastal slopes of subantarctic macquarie island, 1980--1995.
\newblock \emph{Arctic, Antarctic, and Alpine Research}, 34\penalty0 (3):\penalty0 300--307, 2002.

\bibitem[Lagona(2019)]{Lagona_copulas2019}
F.~Lagona.
\newblock Copula-based segmentation of cylindrical time series.
\newblock \emph{Statistics \& Probability Letters}, 144:\penalty0 16--22, 2019.
\newblock ISSN 0167-7152.
\newblock \doi{https://doi.org/10.1016/j.spl.2018.04.011}.
\newblock URL \url{https://www.sciencedirect.com/science/article/pii/S0167715218301652}.
\newblock Advances in statistical methods and applications for Climate change and Environment.

\bibitem[Le~Roux and McGeoch(2008{\natexlab{a}})]{le2008changes}
P.~C. Le~Roux and M.~A. McGeoch.
\newblock Changes in climate extremes, variability and signature on sub-antarctic marion island.
\newblock \emph{Climatic change}, 86\penalty0 (3):\penalty0 309--329, 2008{\natexlab{a}}.

\bibitem[Le~Roux and McGeoch(2008{\natexlab{b}})]{le2008rapid}
P.~C. Le~Roux and M.~A. McGeoch.
\newblock Rapid range expansion and community reorganization in response to warming.
\newblock \emph{Global Change Biology}, 14\penalty0 (12):\penalty0 2950--2962, 2008{\natexlab{b}}.

\bibitem[Le~Roux and McGeoch(2008{\natexlab{c}})]{le2008spatial}
P.~C. Le~Roux and M.~A. McGeoch.
\newblock Spatial variation in plant interactions across a severity gradient in the sub-antarctic.
\newblock \emph{Oecologia}, 155:\penalty0 831--844, 2008{\natexlab{c}}.

\bibitem[Ley and Verdebout(2017)]{Ley_etal2017}
C.~Ley and T.~Verdebout.
\newblock \emph{Modern directional statistics}.
\newblock Chapman and Hall, 2017.

\bibitem[L{\"o}ffler(1984)]{loffler1984macquarie}
E.~L{\"o}ffler.
\newblock Macquarie island: A wind-molded natural landscape in the subantarctic.
\newblock \emph{Polar Geography}, 8\penalty0 (4):\penalty0 267--286, 1984.

\bibitem[Ludwig et~al.(1999)Ludwig, Tongway, and Marsden]{ludwig1999stripes}
J.~A. Ludwig, D.~J. Tongway, and S.~G. Marsden.
\newblock Stripes, strands or stipples: modelling the influence of three landscape banding patterns on resource capture and productivity in semi-arid woodlands, australia.
\newblock \emph{Catena}, 37\penalty0 (1-2):\penalty0 257--273, 1999.

\bibitem[McCune and Keon(2002)]{mccune2002equations}
B.~McCune and D.~Keon.
\newblock Equations for potential annual direct incident radiation and heat load.
\newblock \emph{Journal of vegetation science}, 13\penalty0 (4):\penalty0 603--606, 2002.

\bibitem[McGeoch et~al.(2008)McGeoch, le~Roux, Hugo, and Nyakatya]{mcgeoch2008spatial}
M.~McGeoch, P.~C. le~Roux, E.~A. Hugo, and M.~J. Nyakatya.
\newblock Spatial variation in the terrestrial biotic system.
\newblock In \emph{The Prince Edward Islands: land-sea interactions in a changing ecosystem}, pages 245--276. SUN MeDIA, 2008.

\bibitem[Michalet et~al.(2014)Michalet, Le~Bagousse-Pinguet, Maalouf, and Lortie]{michalet2014two}
R.~Michalet, Y.~Le~Bagousse-Pinguet, J.-P. Maalouf, and C.~J. Lortie.
\newblock Two alternatives to the stress-gradient hypothesis at the edge of life: the collapse of facilitation and the switch from facilitation to competition.
\newblock \emph{Journal of Vegetation Science}, 25\penalty0 (2):\penalty0 609--613, 2014.

\bibitem[Momberg et~al.(2021{\natexlab{a}})Momberg, Hedding, Luoto, and le~Roux]{momberg2021exposing}
M.~Momberg, D.~W. Hedding, M.~Luoto, and P.~C. le~Roux.
\newblock Exposing wind stress as a driver of fine-scale variation in plant communities.
\newblock \emph{Journal of Ecology}, 109\penalty0 (5):\penalty0 2121--2136, 2021{\natexlab{a}}.

\bibitem[Momberg et~al.(2021{\natexlab{b}})Momberg, Hedding, Luoto, and le~Roux]{momberg2021species}
M.~Momberg, D.~W. Hedding, M.~Luoto, and P.~C. le~Roux.
\newblock Species differ in their responses to wind: the underexplored link between species fine-scale occurrences and variation in wind stress.
\newblock \emph{Journal of Vegetation Science}, 32\penalty0 (6):\penalty0 e13093, 2021{\natexlab{b}}.

\bibitem[Morgan et~al.(2010)Morgan, Kirkpatrick, and Di~Folco]{morgan2010wind}
S.~W. Morgan, J.~B. Kirkpatrick, and M.-B. Di~Folco.
\newblock Wind-controlled linear patterning and cyclic succession in tasmanian sphagnum mires.
\newblock \emph{Journal of Ecology}, 98\penalty0 (3):\penalty0 583--591, 2010.

\bibitem[Pewsey and García-Portugués(2021)]{Pewsey_etal2021}
A.~Pewsey and E.~García-Portugués.
\newblock Recent advances in directional statistics.
\newblock \emph{Test}, 30:\penalty0 1–58, 2021.

\bibitem[Rietkerk et~al.(2004)Rietkerk, Dekker, De~Ruiter, and van~de Koppel]{rietkerk2004self}
M.~Rietkerk, S.~C. Dekker, P.~C. De~Ruiter, and J.~van~de Koppel.
\newblock Self-organized patchiness and catastrophic shifts in ecosystems.
\newblock \emph{Science}, 305\penalty0 (5692):\penalty0 1926--1929, 2004.

\bibitem[Rietkerk et~al.(2021)Rietkerk, Bastiaansen, Banerjee, van~de Koppel, Baudena, and Doelman]{rietkerk2021evasion}
M.~Rietkerk, R.~Bastiaansen, S.~Banerjee, J.~van~de Koppel, M.~Baudena, and A.~Doelman.
\newblock Evasion of tipping in complex systems through spatial pattern formation.
\newblock \emph{Science}, 374\penalty0 (6564):\penalty0 eabj0359, 2021.

\bibitem[Selkirk(1998)]{selkirk1998active}
J.~Selkirk.
\newblock Active vegetation-banked terraces on macquarie island.
\newblock \emph{Zeitschrift fur Geomorphologie}, 42\penalty0 (4):\penalty0 483--496, 1998.

\bibitem[Smith and Mucina(2006)]{smith2006vegetation}
V.~R. Smith and L.~Mucina.
\newblock Vegetation of subantarctic marion and prince edward islands.
\newblock \emph{The Vegetation of South Africa, Lesotho and Swaziland. Strelitzia}, 19:\penalty0 698--723, 2006.

\bibitem[Wehrly and Johnson(1980)]{Wehrly_etal1980}
T.~E. Wehrly and R.~A. Johnson.
\newblock Bivariate models for dependence of angular observations and a related markov process.
\newblock \emph{Biometrika}, 67(1):\penalty0 255-- 256, 1980.

\bibitem[Wells and SenGupta(2010)]{wells2010advances}
M.~T. Wells and A.~SenGupta.
\newblock \emph{Advances in Directional and Linear Statistics: A Festschrift for Sreenivasa Rao Jammalamadaka}.
\newblock Springer Science \& Business Media, 2010.

\bibitem[Yedlapalli et~al.(2023)Yedlapalli, Kishore, Boulila, Koubaa, and Mlaiki]{yedlapalli2023toward}
P.~Yedlapalli, G.~N.~V. Kishore, W.~Boulila, A.~Koubaa, and N.~Mlaiki.
\newblock Toward enhanced geological analysis: A novel approach based on transmuted semicircular distribution.
\newblock \emph{Symmetry}, 15\penalty0 (11):\penalty0 2030, 2023.

\bibitem[Zhang et~al.(2017)Zhang, Zhang, Evans, and Huang]{zhang2017vegetation}
F.~Zhang, H.~Zhang, M.~R. Evans, and T.~Huang.
\newblock Vegetation patterns generated by a wind driven sand-vegetation system in arid and semi-arid areas.
\newblock \emph{Ecological Complexity}, 31:\penalty0 21--33, 2017.

\end{thebibliography}

\end{document}